# Effectiveness of Crypto-Transcoding for H.264/AVC and HEVC Video Bit-streams

[1]Rizwan A. Shah, [1,2]Mamoona N. Asghar, [1]Saima Abdullah, [3]Martin Fleury and [4]Neelam Gohar

[1] *Department of Computer Science & IT, The Islamia University of Bahawalpur, Pakistan*

[2]*Software Research Institute, Athlone Institute of Technology, Athlone, Ireland*

[3]*School of Computer Science and Electronic Engineering, Colchester, CO4 3SQ, UK*

[4]*Department of Computer Science, Shaheed Benazir Bhutto Women University, Peshawar, Pakistan*

Correspondence should be addressed to Martin Fleury, fleury.martin55@gmail.com

**Abstract:** To avoid delays arising from a need to decrypt a video prior to transcoding and then re-encrypt it afterwards, this paper assesses a selective encryption (SE) content protection scheme. The scheme is suited to both recent standardized codecs, namely H.264/Advanced Video Coding (AVC) and High Efficiency Video Coding (HEVC). Specifically, the paper outlines a joint crypto-transcoding scheme for secure transrating of a video bitstream. That is to say it generates new video bitrates, possibly as part of an HTTP Adaptive Streaming (HAS) content delivery network. The scheme will reduce the bitrate to one or more lower desired bit-rate without consuming time in the encryption/decryption process, which would be the case when full encryption is used. In addition, the decryption key no longer needs to be exposed at intermediate middleboxes, including when transrating is performed in a cloud datacenter. The effectiveness of the scheme is variously evaluated: by examination of the SE generated visual distortion; by the extent of computational and bitrate overheads; and by choice of cipher when encrypting the selected elements within the bitstream. Results indicate that there remains: a content; quantization level (after transrating of an encrypted video); and codec-type dependency to any distortion introduced. A further recommendation is that the Advanced Encryption Standard (AES) is preferred for SE to lightweight XOR encryption, despite it being taken up elsewhere as a real-time encryption method.

# 1 Introduction

Video content can be streamed to a device for viewing without the need for other than partial storage on the device, though the video stream is stored on a server and possibly encrypted as well. For example, using the most mature form [1] of HTTP Adaptive Streaming (HAS), that is HTTP Live Streaming (HLS), a client device can dynamically select from different representations of the stream according to available bandwidth. Each of these versions might be first transcoded [2] to one of (say) eight versions, with a ninth audio stream. Transcoding might be through changing the spatial resolution, the temporal resolution, or the bitrate, which in effect involves a reduction in the Signal-to-Noise Ratio (SNR), which is commonly interpreted as a reduction in video quality. This paper assesses the effectiveness of changing the bitrate, i.e. transrating, when that bitstream is encrypted. Transcoding may also involve changing the video format, for example between that of one codec to another, though this is usually not required for HAS and is not assessed herein. Prior to transcoding for HLS, the original video stream might be delivered to a Content Delivery Network (CDN) by Real-Time Messaging Protocol (RTMP). Such a CDN, might be part of a CDN as a Service (CDNaaS) [3] cloud-based offering. As remote processing is involved, additional security issues arise if content is not protected. Unfortunately, content protection through encryption results in additional latency at intermediate transcoders due to the need for decryption. The aim of this paper is to avoid that additional latency by allowing the video to be transcoded without decryption taking place. At the same time, the paper seeks to establish whether accelerated forms of encryption/decryption can further reduce the latency, without significantly affecting the content's security. This is a timely study, given the increased prevalence of transcoding in the mobile Internet, which was not around when transcoding was originally applied to statistical multiplexing of broadcast video.

Encryption allows protection of content against illegal access to a video server and also protects each video stream during transport across the Internet. Content protection is deemed necessary to ensure the commercial viability of a video streaming service because otherwise there would be no monetary incentive to make videos and distribute them. For example, the HLS specification [4] supports full encryption of video segments within a representation using Advanced Encryption Standard (AES) key length 128 and the Encrypted Media Extensions (EME) W3C specification describes key management for HTML5 video [5].

RTMPE also supports full encryption through the Rivest Cipher 4 (RC4) stream cipher. As RC4 is now considered cryptographically insecure by the Internet Engineering Task Force (IETF) [6] and, as a result, is not included in browsers such as Internet Explorer 11, alternatively, the RTMP video stream can be wrapped within a Transport Layer Security (TLS) session. However, these forms of full encryption are sometimes called naïve encryption because they do not exploit the features of video, essentially treating it as text. Consequently, it may result in a significant performance overhead, especially if software-only encryption occurs [7].

However, compared to full encryption, with selective encryption (SE) [8] not only is the amount of data encrypted reduced but some forms of SE are decoder format-compatible. If the form of SE is not format compatible, as occurred in [9] for the I-frame encryption option, then the transcoder has to be modified. As transcoder modification reduces the generality of the solution, requiring all transcoders to be modified, this paper only considers decoder compatible forms of SE. A similar observation applies to secured hardware transcoders such as that of the Secure Video Processor (SVP) alliance [10], which may also be costly compared to an unsecured transcoder. Thus, it is possible to ensure the video is unwatchable owing to distortions, while at the same time permit bitrate transcoding without decryption. For example, in the RTMP delivery of the original video to a CDNaaS, were SE to be employed, there would be no intermediate storage of the decryption key, which is only held by the client device. As previously mentioned, the client device also does not store a streamed video, unless deserialization software has illicitly been installed. In addition, the selectively encrypted components can be encrypted by a standardized cipher such as AES operating in a streaming mode such as Cipher Feedback (CFB). After encryption, those encrypted components are normally replaced in the video bitstream in the interests of format compatibility.

Apart from its use in HAS, video content can be further compressed by transcoding to reduce the bit-rate of the video in order to match the capability of a user's device, such as its processing capacity, which will affect the viable display frame rate. Spatial resolution is the other main factor that can reduce the bitrate of the original video content. The resolution of a user's device may be as low as Common Intermediate Format (CIF), as high as 1280 × 720 pixels/frame (High Definition or HD), or even Ultra HD (UHD) resolution [11]. However, though spatial resolution switching does occur in HAS systems, in [12] it was recommended that the spatial resolution of the target device is first adjusted for, after which different quality representations are selected by a client device. In [13], quality switching through encoding was found to be the most common form of representation considered in research. Thus, this paper considers quality transcoding, especially as temporal switching, though effective in terms of the resulting Quality-of-Experience (QoE) has limited impact on the bitrate [12].

Transcoder banks are now common as intermediate devices sitting between mobile devices of various types, such as smartphones, tablets. They provide a way of mediating between high-quality source video, typically held at a server, and the processing capability of the target device, along with any bitrate restrictions on the network path to the device. Should these transcoders or other intermediate devices need to decrypt the video stream then the decryption key is exposed. There is also a key management overhead involved in supplying the key to any intermediate devices, not only to transcoders but devices that might insert logos or watermarks. Another application could be through video transcoding at a satellite and additionally video transcoding already takes place as part of the broadcast statistical multiplexing process

Transcoding of the quality level is also common in digital TV broadcast systems for the purpose of statistical multiplexing TV programs onto the transmission channel. Larger values of the Quantization Parameter (QP) produce a more compressed version of the original video, so that increasing the QP through transcoding can additionally support a 'pay-per-quality' service, as originally described in [14]. Of course, the original compressed video must be encoded with a low QP or high quality, as it is impossible to increase the quality of a video through transcoding. However, for any such 'pay-per-quality' scheme, content protection through encryption is required. Because such a scheme requires decryption before and re-encryption after transcoding, the complexity and transcoding latency will be increased [15]. In this paper, we present a way of transcoding with selective content encryption which aims to reduce those overheads. One should also remark that in the video plus depth 3D video format [16], the depth information is stored as a conventional video, in addition to the normal 2D video. Therefore, the same method of crypto-transcoding could be applied to the depth video stream in the video plus depth format.

Raw video in YUV format is encoded (compressed) prior to encryption. Fig. 1 shows a classical transcoding system, in which the video is fully decrypted before altering the QP and re-encrypting.

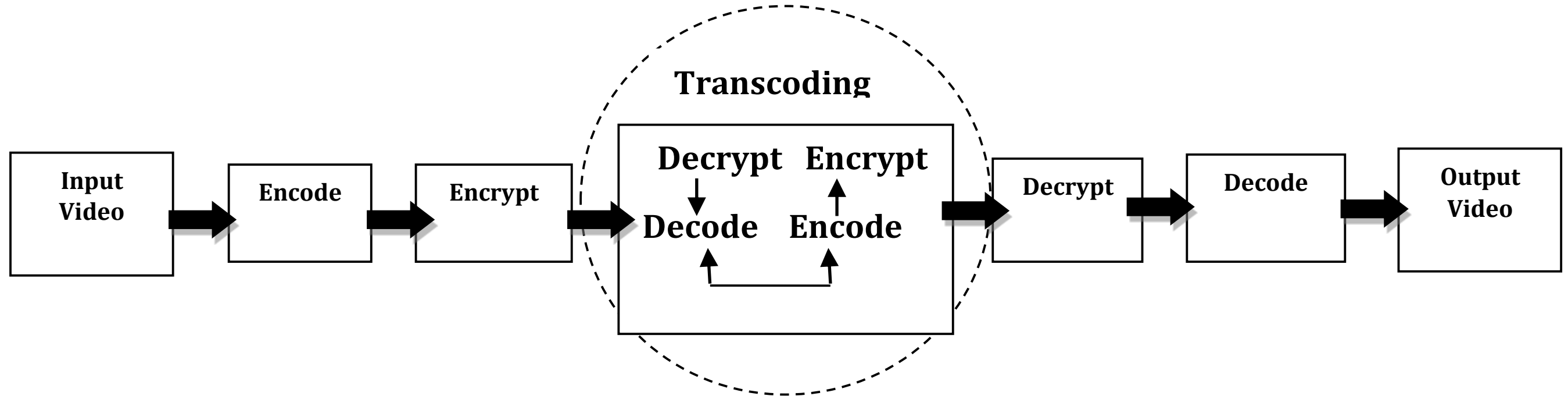

**Fig. 1** Classical transcoding system

Though a number of transcoder designs [17] have sought to transcode without fully decoding the video, in order to decrease latency, some of these designs are vulnerable in one way or another to temporal error drift due to lack of synchronization between the original encoder and the remote decoder. In open-loop transcoder designs, because no reference is made to the frame store before re-quantization or transform coefficient pruning, temporal error drift builds up across a Group-of-Pictures (GoP), even though processing time is reduced. Closed-loop designs seek to approximate the cascaded architecture of Fig. 1 by reintroducing the transform stage; however, though drift is minimized, extra processing latency is re-introduced. Instead, the proposed crypto-transcoder, while avoiding temporal error drift, reduces the time previously taken-up by encryption and decryption. As a result, all processing takes place within the same transcoder block. Apart from reducing processing latency, decrypted video content is no longer exposed to other intermediate devices during a video stream's journey across the Internet. As the focus of this work is on encryption overhead rather than transcoder architectures, the use of alternative transcoder architectures will be the subject of future research. The need for the proposed scheme has increased. Since 2011, according to [18], the personalization phase of video delivery has occurred, by which cloud storage and cloud-based streaming of cached video [19] has become common after suitable representations have been generated by means of transcoder banks. Storage on local video servers in a way is even more vulnerable because professional security management may not be available, implying that encryption is also needed for non-cloud storage.

The key factor of the transcoding system represented in Fig. 1 is that there is a need to decrypt the video before transcoding is started and after transcoding to re-encode before performing re-encryption. As previously mentioned, in Fig. 1's classical system, there is a need to expose the video content during transcoding and the decryption keys are also exposed at the transcoder, as otherwise decoding will fail. In contrast to the classical system of Fig. 1, the crypto-transcoding scheme in this paper has the following research contributions to traditional secure transcoders:

1. The scheme presented in this paper employs a selective-encryption method that works only on uniformly-distributed selected syntax elements of the compressed video stream. Owing to this strategy, a selectively-encrypted video stream is decoder compliant, which means that, despite encryption, the video stream can be transcoded. As a result, there is no need to decrypt and then re-encrypt at the transcoder site.
2. The process of changing the QP otherwise works exactly as before. Finally, the decryption procedure is always performed by a target device and not at any intermediate point, where the content and/or decryption keys may be exposed.
3. SE is applied at the entropy-coding stage of the encoder as a crypto-compression scheme, so that there is limited additional computational overhead from the encryption process.
4. Choosing SE rather than full encryption runs a risk that content may not be sufficiently distorted and, so in some way, expose the video content any way without the need for decryption. Thus, an additional, significant contribution of this paper is to assess the effectiveness of the format compatible and compression-friendly approach to encryption for secure transcoding. The structural distortion analysis of crypto-transrated videos is done through the objective quality metrics in the experiments.
5. For experiments, we have conducted estimates of: the computation involved, the file sizes, and the effect, as part of the selective encryption process, of choosing various syntax elements of a Context Adaptive Binary Arithmetic Coding (CABAC) entropy coder [20]. A part of that evaluation process was the performance of the H.264/Advanced Video Coding (AVC) standard [20] and the more recently standardized High Efficiency Video Coding (HEVC) codec [21]. The intention was not to compare their relative compression performance, which has already been extensively explored. However, much legacy video content remains in H.264/AVC format or even in MPEG-2 codec format. However, it is possible to use format transcoding between MPEG-2 to H.264/AVC coding [22]. Therefore, an important part of this paper's contribution is the implementation of the proposed scheme over both more recent standardized codecs under transcoding.
6. Usually encryption in the context of transcoders is performed through a full-strength cipher, often the Advanced Encryption Standard (AES) cipher, before appropriate replacement in the compressed video stream. Despite its security strength, the AES has complex rounds, which take up much computation. In this paper, two ciphers are tested for crypto-transcoding through selected entropy-coder syntax elements. In experiments, an Exclusive OR (XOR) cipher is tested for both the H.264/AVC and HEVC encoder and then the AES cipher is also utilized in a stream cipher mode over the HEVC encoder. For speed, it is possible that the XOR cipher might be used for that purpose, if the security can be enhanced by means of One Time Pad (OTP) session keys. The current paper, therefore, also considers that alternative too.

The remainder of this paper is organized as follow. Section 2, describes the process of the SE used in this paper, the available transcoder architectures, and other background necessary for an understanding of the paper. Section 3 is a review of related work research in the domain of secure transcoding for HEVC. Then Section 4 outlines the evaluation methodology employed. Subsequently Section 5, considers the effectiveness of crypto-transcoding through experiments to determine the combined effect of SE and transrating according to the visual effect or distortion and the computational and bit-rate overheads, when using either of the two standard codecs. Results taken with two ciphers are also included. A comparative analysis of previous schemes and the current one is also included. Finally, Section 6 draws some conclusions concerning this research.

# 2 Background

This Section provides basic brief introductions to the main components of the scheme. It is not intended to be a comprehensive or full description of these components. In particular, the following Section describes the SE method used in this paper.

## 2.1 Selective encryption

Chaotic-map based full encryption does not have the performance penalties of the naïve encryption schemes described in Section 1. However, if encryption occurs before transcoding takes place, the video must first be decrypted prior to processing, which implies that the decryption key is also exposed at a CDN server. As in the future CDN servers may be placed remotely on a cloud, there is a risk from third-party contractors, who may operate in data-centers outside the legal jurisdiction of the content owner. For live video processing or interactive video processing, the need to decrypt and then re-encrypt will cause a significant delay if full encryption is employed. It is also no longer possible to perform intermediate video processing of video chunks if full encryption has been performed. For example, it is no longer possible to insert logos or watermarks without first decrypting the video stream.

The XOR is widely used in chaotic schemes by the researchers. An interesting study is [23], which considers the use of XOR encryption both in chaotic stream ciphers and two other schemes. Chaotic encryption is designed to avoid the computational overhead of full encryption with a block-based cipher but as the authors of [23] indicate, it has shortcomings if XOR encryption is employed. However, unlike the current paper, SE is not considered. In [24] there is a comparison between: 1) an SE scheme using AES encryption; 2) an SE scheme using AES to generate pseudo-random numbers prior to XOR encryption; and 3) an SE scheme based on chaotic generation of a random stream of numbers prior to XOR encryption. The latter is shown to improve considerably in terms of processing time compared to option 2). However, the originators of scheme 3) do not consider the impact of intermediate transcoding.

Instead, Selective Encryption (SE) [25], as used herein, provides a lightweight procedure for video content confidentiality, as it does not encrypt all video data but selects the most influential or most important syntax elements from the multimedia content (herein video) and then encrypts those elements. Because of the reduction of encrypted material, SE reduce computational overhead compared to full (or its subset naïve) encryption. Thus, this approach lends itself to real-time or interactive applications of video streaming such as video phone, video conferencing, and telemedicine. However, not all types of SE provide efficient or sufficient content protection. Some SE algorithms exhibit weaknesses in terms of: an additional bit-rate overhead; lack of decoder compliance; and insufficient confidentiality. Nonetheless, these weaknesses can be addressed by ensuring that SE is performed at the final stage of a hybrid video encoder [26] i.e. entropy-coding stage, and after ensuring that statistical distribution of encrypted syntax elements will not be altered [27]. Only then does SE become beneficial as no or limited extra bit-rate overhead occurs. Notice also that in [28] the resilience of the SE scheme in this paper has already been checked and analysed in respect to a variety of attacks.

The SE utilized in this paper works at the Context Adaptive Binary Arithmetic Coding (CABAC) form (entropy coding) so that the SE scheme can apply both to the H.264/AVC codec and to HEVC. Notice that, though there are some differences in the way CABAC is performed in HEVC compared to H.264/AVC, H.264/AVC methods of SE can be adapted to those of HEVC by the conversion methods of [29]. The CABAC encoder works on a number of parameters which potentially could be used in the encryption operation, these being the Coded Block Flag; the Motion Vector Differences (MVDs); the Macroblock (MB) types; the Transform Coefficients (TCs); the delta quantization parameters (dQPs); and the numerical signs of TCs and MVDs. However, not all the syntax elements mentioned above provide decoder compliance and so in this paper we select: the signs of TCs and the signs of MVDs, which we abbreviate to MV signs. Due to this selection, the proposed scheme allows SE to take place without decryption and re-encryption when transrating takes place. One key determinant of whether a syntax element is suitable for selection is whether it is by-pass coded, i.e. whether or not it affects the context adaptation models. Elements, such as the above two, are selected because they do not affect the context models. In addition, signs can reasonably be expected to follow a Uniform distribution, before and after encryption. Therefore, in a long-term statistical sense there is no bitrate overhead from this type of SE, even though for particular video streams it may turn out that there is some overhead.

An alternative to decoder-compatible SE exists, which might permit transcoding without prior decryption. That is video can be encrypted prior to compression. However, unless specialist forms of encryption are deployed, encryption removes the correlation that compression exploits, resulting in a loss of compression efficiency. Permutation-based forms of encryption can preserve or even enhance the correlation within a video frame. On the other hand, in [30], which described its proposed security as ‘reasonable’, the method was confined to spatial-only codecs. In fact, other encryption-then-compression schemes such as [31], though they offer a solution to a need for intermediate processing, including transcoding, appear confined to spatially-coded images and may have weak compression performance [32].

## 2.2 Ciphers: Advanced Encryption Standard (AES) and real-time XOR

The symmetric encryption cipher AES [33] was chosen as one encryption option for the SE elements. AES has low memory requirements and has been designed to guard against timing attacks. The AES structure is shown in Fig. 2. The encryption procedure uses a set of especially derivative keys called round keys. The following AES steps are for the encryption of a 128-bit block:

1. Calculate a set of round keys from the cipher key.
2. Arrange a state array with the plaintext block data.
3. Add the initial round key to the preliminary state array.
4. Execute nine rounds of state operations.
5. Perform the tenth and last round of the state operation.
6. Duplicate the final state array as the outputted encrypted ciphertext.

As AES employs a single key with a limited key length, the efficiency is increased in terms of computational time, and memory consumption compared to asymmetric-key algorithms. Additionally, compared to prior standardized symmetric ciphers, AES provides a shield against many attacks [34].

The logical function XOR can optionally be applied as a symmetric cipher to the selected CABAC binary bins aggregated into 128-bit blocks. Its one-step operation results in rapid encryption. However, especially if the same key is repeatedly used, the security is weak, being vulnerable to a known-plaintext attack by XORing the plaintext with the ciphertext to output the key. The cipher is also vulnerable to flipping of the cipher text so that a valid but incorrect 'message' is generated. This effect is termed malleability, when decryption takes place. However, by means of a continually changing key generated with a Pseudo-Random Number Generator (PRNG), the confidentiality is greatly enhanced, if the initial seed can be securely distributed and the PRNG is 'sufficiently random'. Another possibility is to establish a long-term AES key through the Diffie-Helman key distribution protocol [35]. Thereafter, a session key, possibly for each video frame depending on the level of protection required, is then AES encrypted and included in the header of the XOR SE file. This form of lightweight encryption is also gaining prominence in lightweight encryption for smart grid applications, for example [36].

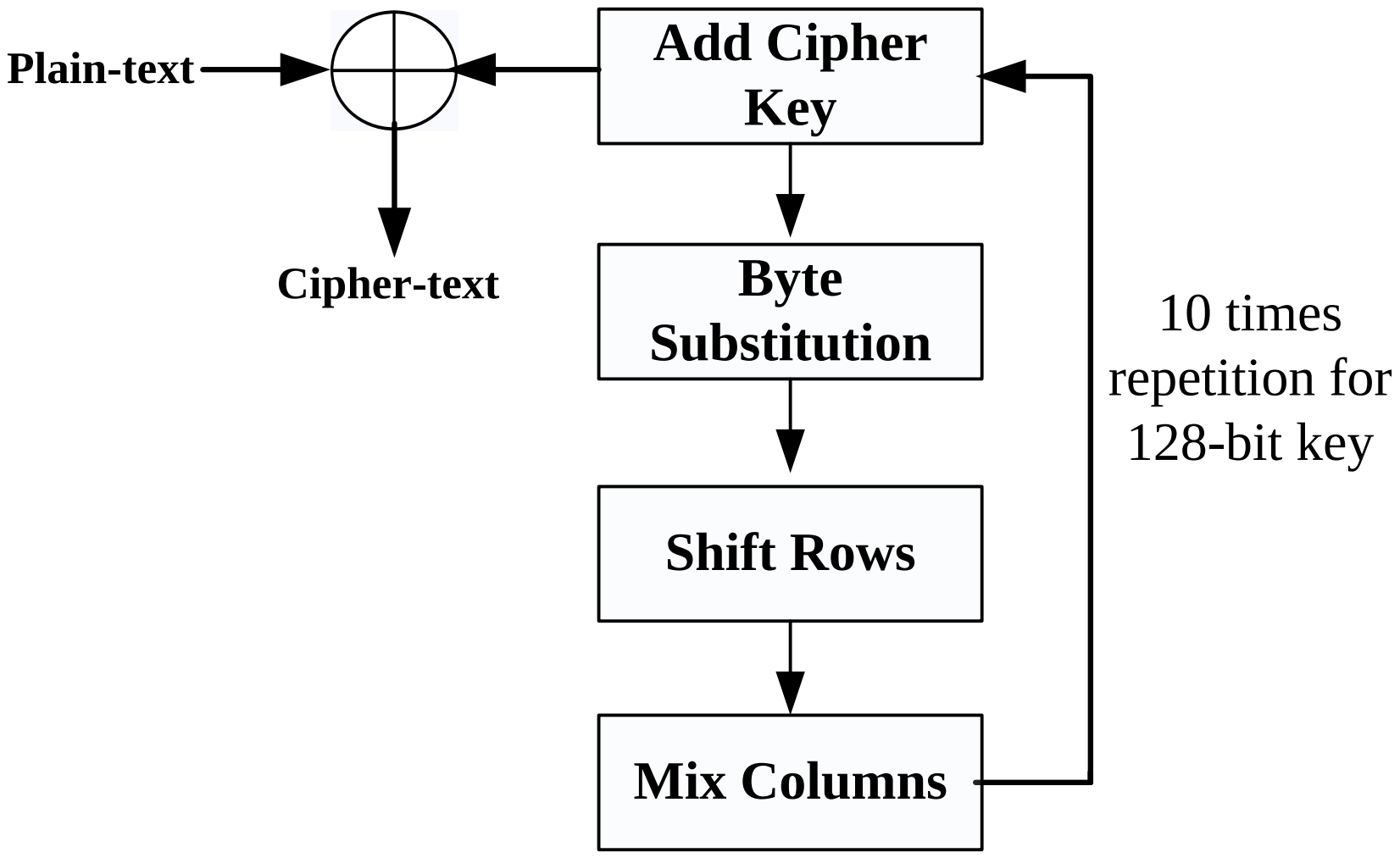

**Fig. 2** AES structure to encrypt 128-bit plaintext block

## 2.3 Transcoding

Video transcoding is the process of converting a compressed video from one form into another. Transcoding is performed on the basis of parameters such as the bitrate, frame rate, and spatial resolution. It is possible to convert from one codec standard to another, such as when a video encoded in a newer format, for example H.264/AVC, commonly employed for video over wireless, is converted to a legacy format such as MPEG-2, so that it can be broadcast over digital TV networks [37]. One of the main uses of transcoding continues to be reducing the bit-rate of a pre-compressed video stream according to the available channel bandwidth. Increasingly owing to the proliferation of networked devices, different clients may use different ways to access the Internet [38]. Each access network type has different channel characteristics such as bandwidths, bit rate errors and packet loss rates. At the user end, different networked devices including smartphones and desktop PCs are used for browsing the Internet. All end-user devices, including Set-Top Boxes, vary in terms of resources such as computing power and display resolution. To deliver video streaming data to users connected with different types of networks and having different terminals may need to be adapted dynamically at intermediate locations within a network [39]. Transcoding at intermediate relays has similar security implications to that when transporting encrypted source video in preparation for HAS representation generation within CDNs, described in Section 1. Encrypted video is also transcoded at various points during in-house production of films or TV programs, when the need for decryption and re-encryption is a considerable problem.

Transcoding is one way to handle and accomplish the conversion task, the other way being scalable video. However, there may be a bitrate overhead arising from scalable coding, though HEVC variants of SVC [40] have gone some way to address that issue by considering inter-layer prediction. Encryption of H.264/SVC was considered in [27]. It is also possible [41] to apply a form of transcoding to encrypted scalable video after packetization. This works by employing the unencrypted packet headers to guide truncation of the encrypted packet payload. However, further consideration of transcoding of scalable video is beyond the scope of the current paper.2.4 Transcoding Architectures

There are several types of transrating architectures [17] [42] [43]. The relative advantages/disadvantages of these architectures are summarized in Table 1.The decoding-encoding cascaded architecture is the classical architecture, already described in Section1.

**Table 1.** Comparison among different transcoding architectures

| Parameters | Cascaded Decoding-Encoding | Open Loop | Closed Loop |
|---|---|---|---|
| Computational cost | This architecture is more costly in terms of computation overhead and processing units. | The open loop architecture is the fastest and the simplest means of video trans-coding. | This architecture approximates the cascaded decoding-encoding architecture by reintroducing a transform stage. |
| Error signal | | In an open loop architecture, the output is neither measured nor fed-back for comparison with the input. | In a closed-loop architecture, an error signal is fed-back to minimize the temporal error drift. |
| Reference frames | A reference frame is used to minimize the difference between the input and output frame. | Reference frames from a frame buffer are not used in the processing. | For each reference frame, feedback (the difference between the actual frame and desired frame) is used to take corrective action. |
| Video drift | A reference frame is stored in a decoded frame buffer, which is utilized properly to remove temporal drift. | In this architecture temporal error drift is increased, particularly if high-frequency DCT coefficients are removed from the residual information. | Temporal drift is removed owing to the use of a reference frame and taking motion compensation as a linear function. |
| DCT/IDCT | Two pairs of DCT/IDCT are used. | The DCT/IDCT block is not used in this architecture. | A single pair of DCT/IDCT is used. |
| Best usage | This architecture can be used as a benchmark transcoder. | Its use straightforward and works best for intra-coded pictures. Latency is minimized. | This architecture is the best compromise for the video transcoding process, though additional latency occurs due to extra latency. |

## 2.4 Standardized Video Codecs

As mentioned in Section 1, the scheme is suitable for both current video codec standards, the H.264/AVC standard [20] and the HEVC codec [21], the differences between which are summarized in Table 2, which is a modified version of that in [44]. Both of these codecs are standardized according to the bitstream format delivered to the decoder. As such they are suitable for use in consumer electronics devices. The HEVC standard is specialized towards HD and even UHD resolution video. As such it has introduced many coding refinements to achieve the required compression ratios to accommodate video streams in those formats across reduced bandwidth links. Though, the compression ratio can be increased by up to 50% by means of an HEVC codec that advantage comes with an increase in codec complexity, which in turn increases processing latency.

**Table 2** Comparison between H. 264/AVC and HEVC standards after [44]

| Category | H.264/AVC | H.265/HEVC |
|---|---|---|
| Names | MPEG 4 or H.264/AVC (Standardized in 2003) | MPEG-H, HEVC or H.265 (Accepted in Jan. 2013) |
| Key Improvement | • Designed to work with HD video stream delivery for online and Broadcast<br>• 40-50% reduction of the bit rate compared to prior standards | • Aimed to deal with UHD, 4k, 2k for online and broadcast<br>• 40-50% reduction of the bit rate at the same visual quality matched to previous standard (H.264/AVC) |
| Compression Model | Hybrid spatial-temporal prediction model<br>• Flexible partition of Macro Block (MB), sub-Macro Block for MV prediction<br>• 9 directional modes for intra prediction<br>• 16×16 maximum Macroblock structure<br>• Entropy coding by CABAC or lower complexity Context Adaptive Variable Length Coding (CAVLC)<br>• ½ or ¼ pixel interpolation | Extended hybrid spatial-temporal prediction model<br>• Introduced Coding Tree Units (CTUs) (Coding, Prediction and Transform Units (CU, PU and TU respectively) in quad-tree structure<br>• 35 directional modes for intra prediction<br>• Parallel processing architecture, enhancements in multi-view coding extension<br>• CTU supporting larger block structure (64×64 pixels) with more variable sub-partition structures<br>• Entropy coding is only Context Adaptive Binary Arithmetic Coding (CABAC) |
| Specification | • Support up to 4K (4096 × 2304 pixels/frame)<br>• Supports up to 59.94 fps<br>• 21 profiles; 17 levels | • Support up to 8K UHDTV (8192 × 4320 pixels/frame)<br>• Supports up to 300 fps<br>• 3 approved profiles, draft for additional 5; 13 levels |
| Drawback | Unrealistic for UHD content delivery owing to high bitrate requirements.<br>Low frame rates unsuitable for higher resolutions. | Computationally expensive (~300% +) due to larger PUs and expensive Motion Estimation (intra prediction with more modes, asymmetric partitions in inter prediction, 1/8$^{th}$ pixel interpolation) |

# 3 Related Work

This Section considers the SE and possible transcoding of HEVC, as prior SE of H.264/AVC has been already considered in surveys such as [8] [45]. The Section additionally considers recent lightweight encryption schemes, as these have a bearing upon the investigation of XOR encryption of video in this paper.

The authors of [29] discussed how their prior H.264/AVC SE scheme, applied at the entropy coding stage to CABAC binstrings, could be adapted to HEVC, even though a somewhat different form of element coding (truncated Rice code instead of unary code) is employed in HEVC. In the HEVC version also, if real-time AES encryption was to be achieved, there was a need to concatenate elements to be encrypted so that their concatenated length was a power of two. In the prior H.264/AVC version as in the HEVC version, chosen elements of CABAC binstrings were encrypted with AES in CFB mode. Principally, bits of the quantized transform coefficients (QTCs) and motion vector difference (MVDs) were encrypted. By careful choice of the encrypted CABAC parameters, it is possible to make the bitstream format compliant with limited or no impact on the bitstream size, as context modelling is not affected. As encryption does not extend over HEVC's entropy coding slices, potential parallel computing is also not affected. However, the method of [29] is reported by the authors to not be robust to compression domain processing, which includes some forms of transcoding. A minor issue that possibly could be rectified is that an update of the HEVC standard has meant that the method is no longer format compliant.

Hofbauer et al. [46] investigated transparent encryption of HEVC bitstream. Transparent encryption is a form of SE in which the viewer is able to partially view the content with a view to encouraging purchasing of a full quality version. The method worked by flipping the signs of AC transform coefficients signs. The percentage of bit flipping can be varied according to the desired level of transparency. Mid-range video quality, determined by QP, was evaluated. In fact, the authors conceded that the approach is unsuitable for high-quality video because, in this case, some blocks may not actually be transformed, resulting in no bits to flip.

To address issues with pioneering SE of HEVC CABAC elements, in [47], a somewhat different choice of coding elements was chosen, namely coeff_sign_flag, mvd_sign_flag, cu_qp_delta_abs and the suffix of abs_mvd_minus2. As before these bits are extracted, encrypted with AES after concatenation before placing the encrypted bits back in their original positions in the HEVC output bitstream. Though [47] demonstrates resilience against a replacement attacks, i.e. replacing bits known to be encrypted with other bits, and some other threats, within [47] the few pages available did not permit a full cryptanalysis. Preliminary analysis, did however suggest low computational overhead. Recent work [48] has also considered ways to protect the privacy information of individuals. One-way anonymity can be preserved is through a group signature mechanism and these mechanisms can be made flexible from a server's perspective. In [48], they are also made more flexible from the user's perspective.

The authors of [49] proposed an entropy-coding stage SE scheme that avoided affecting syntax elements which potentially might be manipulated by encryption. For example: splicing of video is suggests that Network Abstraction Layer (NAL) headers cannot be encrypted; motion information bits affect no-reference quality assessment [50]; compression domain insertion of watermarks [51] may be affected; and more specifically to the current paper, certain forms of compression domain transcoding [52] can be impacted. In fact, the possibility of combining more than one compression domain process, such as transcoding, watermarking, and encryption needs to be considered [51] [53]. In [49], no choice of syntax elements for SE is made but the trade-off between increase in bitrate for the same video quality and increase in confidentiality or security in general is analysed. For example, the impact of encrypting Sample Adaptive Offset (SAO) filtering parameters has very limited effect on the bitrate but can be thwarted if this form of in-loop filtering is turned off at the encoder.

In [54], the implementation of a symmetric transcoder which incorporates SE and is suitable for smartphones. By symmetric is meant that the transcoder will encrypt video output from an encoder or decrypt video arriving in encrypted form, similar to a cascaded transcoder. By selecting syntax elements that bypass context modelling at the entropy stage, as also used by prior SE schemes, the scheme avoids an increase in the bitrate even though SE has been applied. Because confidentiality may be weakened by only choosing bypass CABAC syntax elements in [55], the authors analyse the trade-offs if non-bypass (regular mode) elements are chosen. By choosing to encrypt the intra prediction modes used, the confidentiality is significantly improved. As in [54] ciphering is embedded in the transcoder rather than the encoder to avoid the need to make SE encoder dependent. In a similar way, the decoder is made independent of decryption, which is performed in the transcoder. However, this appears to leave the video exposed if the symmetric transcoder is place in an intermediate network middlebox. Recently, [56] provides another analysis of CABAC syntax elements suitable for encryption, choosing the coeff_abs_level_remaining element. However, it is unclear what the impact on decoder format compatibility or bitrate overhead is or whether other elements could also be selected.

HEVC tiles allow a video sequence to be decomposed into autonomous rectangular areas. In the SE option of [57], the authors examined Region-of-Interest (RoI) encryption of tiles. It was found that this implied that motion vectors from non-encrypted tiles should not reference encrypted tiles. The result was some loss of rate-distortion performance to achieve RoI SE. However, propagation of encryption outside encrypted tiles was avoided.

Thomas al. [58] considered various secure transcoder systems according to the aims of the system. For example, if the transcoder only works on inter-coded frames (P- and B-frames) in the interests of accelerating transcoding, If full encryption of I-frames occurs then the Cascaded Pixel Domain Transcoder (CPDT) form of transcoding is handicapped by the lack of I-frame data, which the authors address by modifying the CPDT transcoder. SE, rather than full encryption, may also be performed only on intra-coded frames. However if traditional sign-bit encryption is performed then error drift through lack of synchronization between encoder and decoder can occur. Again the authors of [58] suggest a modified sign-bit encryption scheme to restore synchronization.

In [59], new ways of transcoding in the face of HEVC's quad-tree block structures are considered. Basically the proposed solution focuses on reducing the coding options employed. An implemented transcoder resulted in 80% less time for the transcoding process as compared to a conventional cascaded encoder decoder, see Fig. 1, but the coding performance reduced by up to 5%. In [60], computational scalable video transrating was investigated because of the high computation cost of HEVC cascaded transcoding. In this technique, information from the input bitstream prior to transrating was used to reduce the computation after bitrate scaling. This was done by altering the CU and PU structures so that the number of RD evaluations was reduced. Two methods of reducing subsequent computation were considered, namely top to bottom (T2B), bottom to top (B2T) processing of the CU structure. Machine learning techniques aided in this process. Furthermore, PUs were also reduced in number by manipulating information from the input stream.

Because with the advent of HEVC, H.264/AVC standard compressed content may need to be converted to the new standard and in [61] [62] fast transcoders have emerged. However, because this paper is concerned with transrating rather than format transcoding, these transcoder designs are not considered further herein. Because also there is concern over the computation overhead from using HEVC standardized codecs, there have been numerous proposals as to how to reduce that overhead, such as [63]–[68]. However, these are outside the scope of the present work.

In the view of the authors of [69], conventional encryption methods, such as full encryption by AES, are unsuitable for video data because of the computational overhead, even though major encryption providers such as Verimatrix & Widevine use AES encryption for satellite systems. The approach of [69] was to design a wavelet-based hardware codec that is made amenable to encryption, though in commercial usage, the lack of standardization. In [70], a hardware-based, cellular automata method of encryption was proposed. Because the application was video surveillance, RoIs can be selected, reducing the computational overhead further. However, possible loss of compression efficiency caused by the need to treat encrypted RoIs separately and the risk from applying encryption to a limited area does not seem to have been checked. Nonetheless, the method is suitable for lightweight encryption of video collected in an Internet of Things (IoT) video surveillance setting and presented an alternative, lightweight method to the SE of this paper. In the context of IoT-based video surveillance, in [71] only key frames were encrypted with a chaotic encryption scheme. However, the weakness of some tests used to justify the security of such chaotic encryption schemes was questioned in [23], as was the relative encryption speed compared to conventional block-based encryption methods. Instead, the authors of [72], for a secure but lightweight IoT-based video streaming scheme, preferred to employ conventional block-based encryption. However, they reduced the message overhead through a lightweight transport-layer protocol into which they inserted a Hash-based Message Authentication Code (HMAC) authentication field. All the same, SE, as presented in the current paper, remains an effective lightweight encryption method, with advantages of lightweight chaos-based encryption. As developed in the current paper, it also supports intermediate processing, such as the transrating analysed herein.

# 4 Methodology

To remove error drift from the video stream, transcoding was by means of the closed-loop architecture of Table 1. Fig. 3 is a block diagram of the closed-loop architecture, i.e. presented for a generic transrater. We have proposed and implemented the crypto-transcoder in a closed-loop architecture. Therefore, it is worthwhile to explain this architecture here for newcomers to the field. The input bitstream, $R_{in}$, is first variable-length decoded (vld), allowing the embedded MB motion vectors (MVs) to be extracted. Inverse quantization, $q^{-1}$, of the residual transform coefficients then takes place using a QP, $q_1$, originally used by the encoder and placed in the compressed bitstream. A control, *ctrl*, places a desired new QP, $q_2$, ready for re-quantization before placing the residual coefficients within the output bitstream, $R_{out}$. The output bitstream is produced by variable length coding, *vlc*, that is using some form of entropy coding such as CABAC or CAVLC. However, prior to output of the bitsream, the process of motion compensation (*MC)*, takes place based on the stored reference frame(s), which are adjusted for the changed QP. To do this requires spatial-frequency transforming motion-compensated MBs to allow partial synchronization at the decoder for the effect of transcoding at a different QP to the one originally used at the encoder. Thus the motion-compensation loop of Fig. 3, is the closed loop that gives this type of transcoder its name. Also shown in Fig. 3 is the path taken at an end decoder to retrieve the video. Because the computation involved in a closed-loop architecture transcoder is already substantial, it is unwise to increase the computation further by requiring decryption and re-encryption. From Fig. 3, it is apparent, that if multiple representations at different bitrates are to be generated, (say) for HAS, then motion information and de-quantized information from the decoder can be extracted once and those inputs to the encoder can be repeatedly used in the encoder, thus saving on computation. It should also be noticed that if the source bitstream was not already encrypted, e.g. not part of an encrypted bitstream delivered to a middlebox or cloud data-center, then the output could be selectively encrypted at the Variable Length Coding (VLC) stage of Fig. 3. In Fig. 3 DCT is used for Discrete Cosine Transform and IDCT is its inverse.

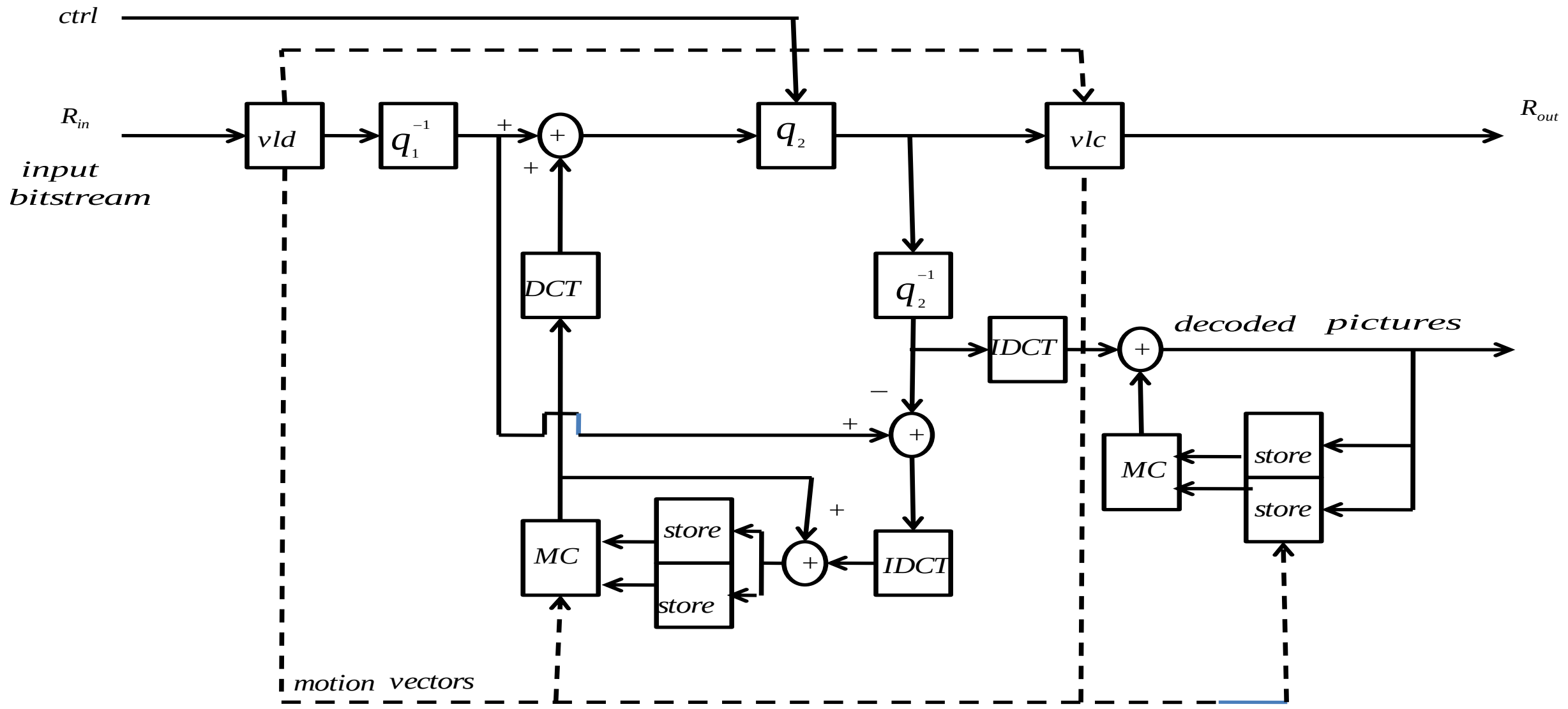

**Fig. 3** Closed loop architecture of transcoders

During motion compensation, typically, each 16 × 16-pixel MB is split up into 4 × 4 pixel blocks in H.264/AVC, while into 8 × 8 pixel blocks in HEVC. (In some cases in HEVC, an 8 × 8 pixel block is further split into 4 × 8 and 8 × 4 pixel block sizes as well.) After the decoding of each frame (and before moving on to the next frame) the Motion Vector (MV) parameters, being the vector components ($MV_x$, $MV_y$) and a pointer(s) to the reference frame (s) used, are extracted. These parameters define the motion compensation for each smallest possible block. To encode and encrypt the bitstream at a number of different QP levels, the stepwise procedure of the proposed scheme is described below and in Fig. 4, for the example case of crypto-transcoding to a set of QPs. The adopted algorithm shown in Fig. 4 assumes that the video is initially encoded at QP = 12, and then transcoded to a set of lower quality video streams, with QP = 24, 36, and 48, given that the range of QPs for H.264/AVC and HEVC is 0–51. In Fig. 4 ME represents motion estimation.

**Step 1:** Encode the raw video with the proposed H.264/HEVC crypto-entropy coder (modified CABAC) along with quantization at the initial QP value, herein QP = 12. Encrypt the selected parameters, i.e. signs of MVD and signs of residual texture information (TCs) of the outgoing H.264/HEVC bitstream.

**Step 2:** In the H.264/HEVC decoder, for every compressed H.264/HEVC encoded bitstream, the horizontal MVx and vertical MVy, along with a pointer(s) to the reference frame(s) used, are extracted from the smallest possible block of the H.264/HEVC bitstream and stored separately in a file. The decoding process is performed without decryption of any parameters and, thus, without the need to supply a key.

**Step 3:** Perform crypto-transcoding with the new QP value but without decryption of the video bitstream, according to the closed-loop architecture of Fig. 3. If necessary, repeat transcoding with one or more different QPs with the same extracted MV parameters from step 2.

**Step 4:** At the receiving end, the received video sequences after decryption and decoding will be able to be watched. Decoding with decryption follows the lower path shown in Fig. 4. (The optional verification step is added for clarity of understanding the whole scheme.)

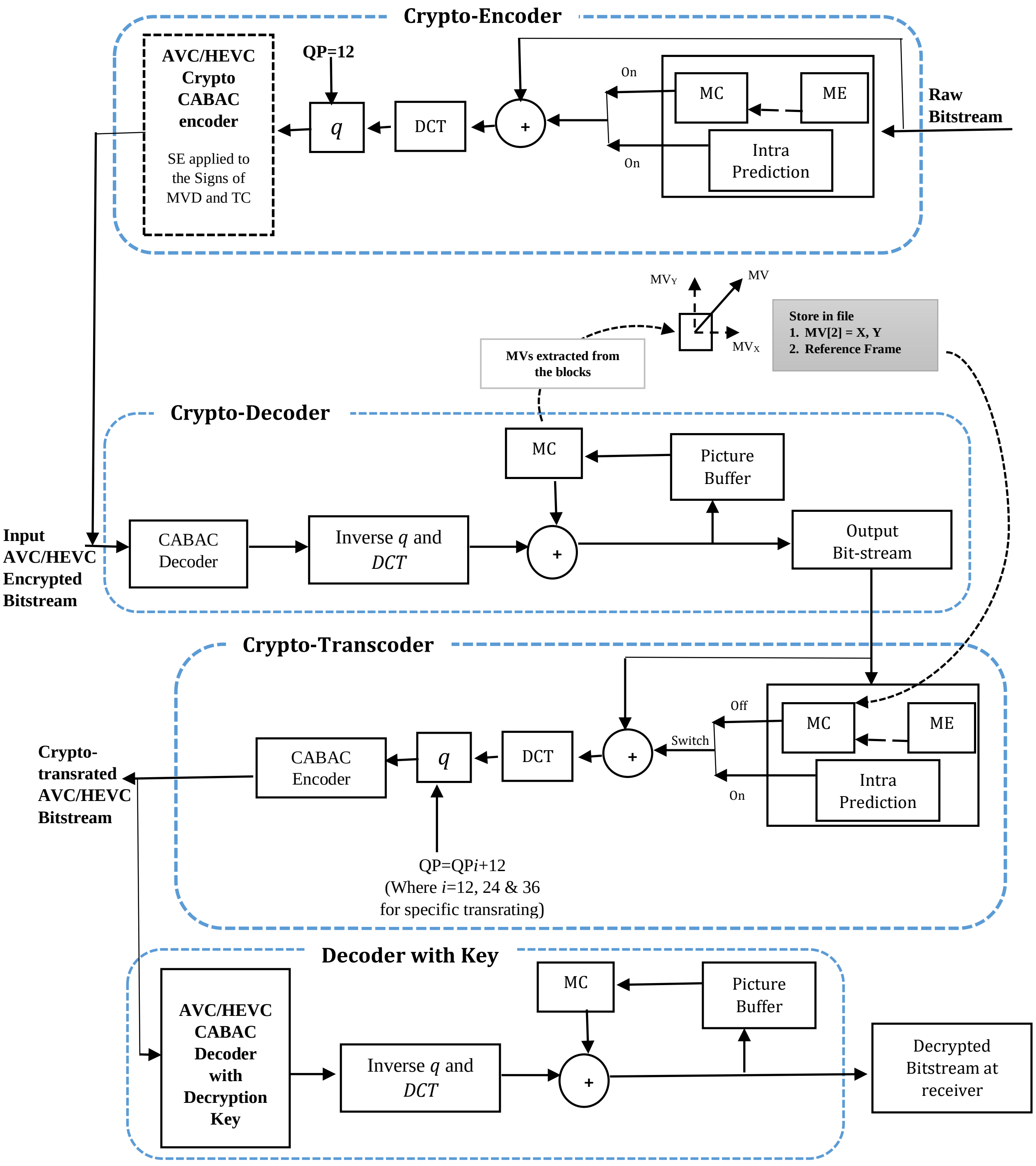

**Fig. 4** Stepwise flow diagram of the proposed crypto-transcoding scheme

# 5 Evaluation

To perform experiments, H.264/AVC reference software JM 18.6 [73] and HEVC reference software HM-15.0 [74] were selected. The machine used for all the experiments was an Intel Core i3 Core 2 Duo (2.10 GHz) processor, with 6 GB RAM and Microsoft Windows 8.1 Professional installed as the operating system. Transcoding on this machine was conducted with the closed-loop architecture of presented in Fig 4. As described in Section 1, video content for HAS can be transcoded into a number of qualities, which in turn determine the bitrate of the different representations available to a client by selecting from a manifest file. The quantization parameter (QP) normally determines the extent of compression (necessary to make bandwidth consumption manageable), which also impacts on the processing required at the target device. This assumes that Variable Bit Rate (VBR) video content is presented to the transcoder, with the result that constant quality representations result. Constant Bit Rate (CBR) video, although it allows video storage to be planned has a disadvantage during transmission because if the content, for a given quality, does not require its bitrate then bandwidth wastage occurs. The well-known FFmpeg software, depending on the underlying encoder selected, allows constrained VBR to be output [75], avoiding a risk of a video stream temporally exceeding its average bitrate. However, in this paper for simplicity in making quality comparisons, VBR is assumed.

Two reference CIF (352 × 288 pixels/frame) video sequences, i.e. *Stefan* and *Mobile* (available from http://www2.tkn.tu-berlin.de/research/evalvid/cif.html) and an HD 720 (1280 × 720 pixels/frame) video sequence, i.e. *Four People*, were transcoded to different QPs (12, 24, 36 and 48). All tested videos are encoded with I, P and B frames with GOP of 16. A QP of 12 results in broadcast quality video, while if the QP is set to 48 the video is very compressed but equally the visual quality is very low in both an H.264/AVC and HEVC codec. Otherwise, the video configuration settings of the original test videos were retained.

Fig. 5 illustrates frames from the test video sequences with calculated average Peak Signal to Noise Ratio (PSNR), an objective measure of video quality measured in decibels (dBs) [76], and average Structural Similarity (SSIM) index [77], which aims to capture the human perceptual response on a scale (usually) of 0 to 1.

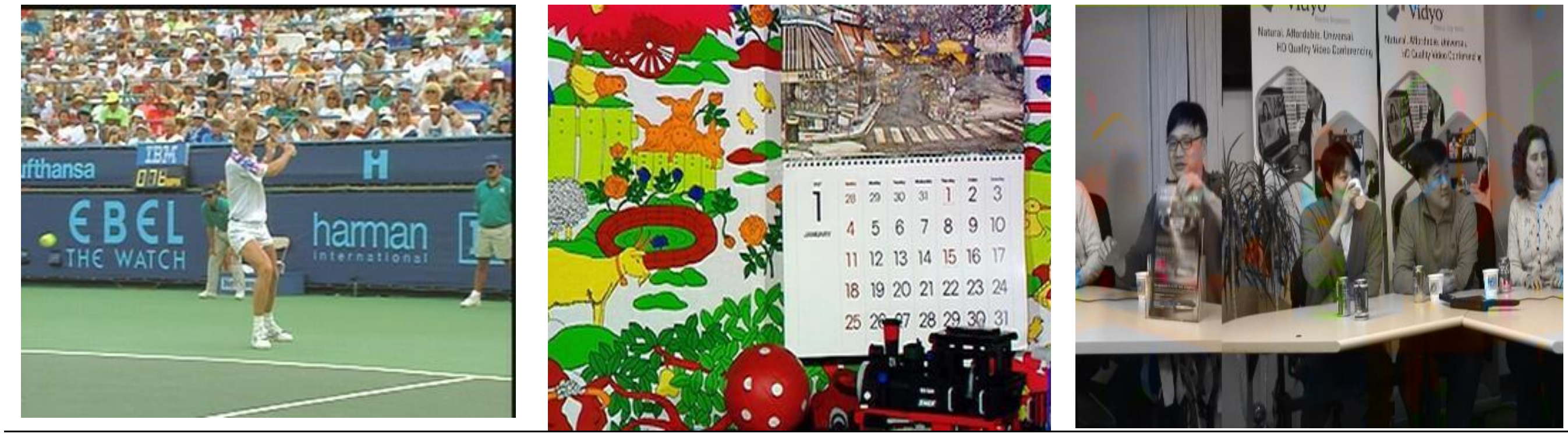

| Stefan Video (frame # 13)<br>PSNR [Y=27.9, U= 46.8, V=41.9] dB,<br>SSIM = 0.9626 | *Mobile* Video (frame # 54)<br>PSNR Y=27.7, U= 40.5, V=37.5] dB,<br>SSIM = 0.9446 | Four People (HD 720 video) (frame # 53)<br>PSNR [Y= 27.7 , U= 46.6 ,V= 54.21] dB,<br>SSIM = 0.9446 |
|---|---|---|

**Fig. 5** Videos frames from the test videos (without crypto-transrating) showing PSNR and SSIM values for the luminance, Y, and chrominance, U and V, components

The above video sequences were firstly compressed by using lossless compression, by setting QP = 0, using either H.264/AVC or HEVC codecs. SE was applied during compression by selecting the TC and MV signs for AES encryption. Subsequently, the compressed versions of the sequences were transrated to different QP levels.

The following equations were used to calculate the various quantities reported. For the bitrate reduction in going from H.264/AVC to HEVC:

$$\Delta\,Bitrate = \frac{(Bitrate)_{HEVC} - (Bitrate)_{AVC}}{(Bitrate)_{AVC}} \times 100 \qquad (1)$$

For visual evaluation after SE, the PSNR and SSIM index value were returned by equations (2) and (4) respectively:

$$\mathrm{PSNR} = 10.\log_{10}\frac{(2^{x}-1)^2}{MSE} \qquad (2)$$

where MSE is the Mean Square Error between the reference video and the video of interest, with x being the bits per pixel, herein being eight to allow comparison between the two codecs.

The PSNR difference was calculated by equation (3):

$$\Delta\, PSNR\ (Y) = \frac{(PSNR_y)_{HEVC} - (PSNRy)_{AVC}}{(PSNy)_{AVC}} \times 100 \qquad (3)$$

and SSIM was calculated by equation (4):

$$SSIM(a,b) = \frac{(2\mu_a\mu_b + c1)(2\sigma_{ab} + c2)}{(\mu_a^2\mu_b^2 + c1)(\sigma_a^2 + \sigma_b^2 + c2)} \qquad (4)$$

where a, b are the two video frames being compared, with $\mu_a$, $\mu_b$being average pixel value (intensity) within a, b respectively, with $\sigma_a, \sigma_b$ being the variance within a,b respectively, and $\sigma_{ab}$ being the covariance. Two variables, c1 and c2, are introduced to stabilize division with relatively small denominators [77].

Encoding time difference was found by equation (5):

$$\Delta\, Enc.Time = \frac{(Enc.\ Time)_{HEVC} - (Enc.\ Time)_{AVC}}{(Enc.\ Time)_{AVC}} \times 100 \qquad (5)$$

Apart from the comparison of video frame distortion by one of the two video quality metrics, PSNR and SSIM, two other processing steps were taken to determine the extent (or lack of distortion). The first of these was by edge-detection using a Laplacian filter [78] and the second of these was by pixelation [79], i.e. lowering of the resolution, of the frames. For both of these effects, an 8 × 8 pixel filter was applied to a video frame to obtain the effects.

## 5.1 Crypto-transcoding with AVC and HEVC

The results of crypto-encoding and then crypto-transrating according to the choice of codec and QP are shown and discussed in this Section. The results are taken according to the steps of the algorithm presented in the Methodology section (Section 4).

From Figs. 6 and 7, the reader will see that the content of all sample video frames is distorted to a greater or lesser extent after applying SE (Fig. 6, 7 (a1, a2)) and then transrated shown in Figs. 6 and 7 (b1, b2, c1, c2, d1 and d2). However, there are conspicuous visual differences between the outputs of the both codecs. In addition, some portions of the frames remain still visible, such as the static calendar in *Mobile* and the tennis court in *Stefan*. This is not surprising as the calendar and tennis court are largely static, whereas the method of SE works on changing values of the MV signs. If there is little motion the MVD will be small and the impact of encrypting the signs will be small. Examining the best quality at QP = 12 and the worst at QP = 48 of the sample *Mobile* frame, it is evident that the impact of SE along with transrating is greater on lower quality video. The same is also particularly apparent for the same QP levels of *Stefan*, especially when the original encoding was with the HEVC codec. This effect may be related to the TC sign encryption. At higher QPs, more of the TCs are reduced to zero and, hence, do not appear in the input to the CABAC engine. Therefore, the effect of flipping the signs of the fewer TCs that are present may have a greater impact. To verify the effectiveness of the crypto-encoding along with transrating over tested videos, video structural distortion analysis is done with Laplacian edge detection or with the pixelate effect, as is demonstrated in Figs. 6 and 7. Notice that a recent way of testing the effectiveness of encryption is to apply edge detection to the distorted frame [80] and pixelation also serves to check the effectiveness of the encryption.

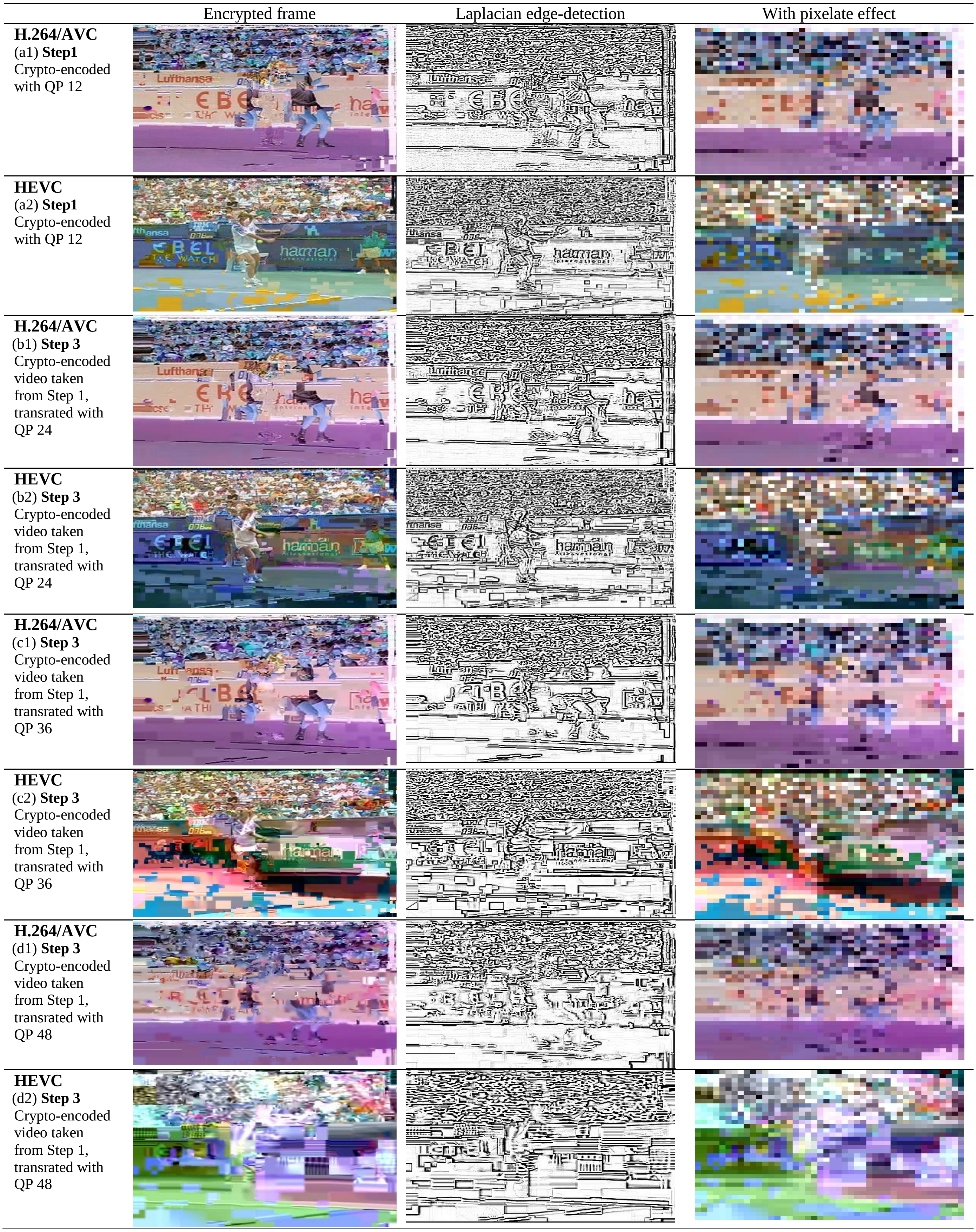

**Fig. 6** Visual results for *Stefan* (frame # 13) after crypt-transrating with different QPs, with edge-detection and pixelate tests

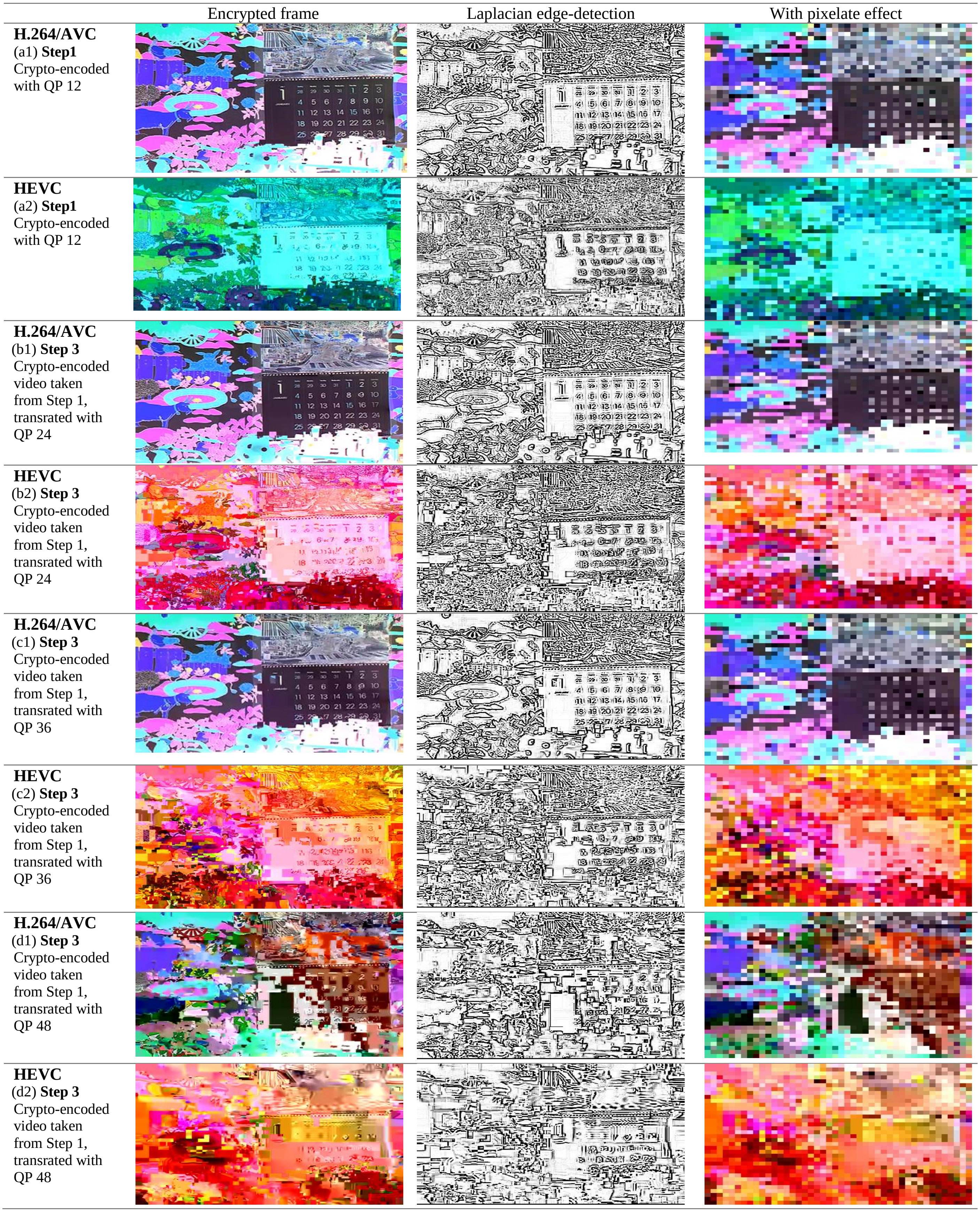

**Fig. 7** Visual results for *Mobile* (frame # 54) after crypto-transrating with different QPs, with edge-detection and pixelate tests

In Tables 3 and 4, it is proved from the output file sizes, HEVC encryption and subsequent transcoding generally results in greater compression at all QP levels. However, this comes at a cost in a considerable time spent in transrating of the encrypted content when HEVC is employed. From the two Tables, in PSNR terms, HEVC SE and subsequent transrating results in less distortion than when H.264/AVC is used. Recall that Tables 3 and 4's PSNR results are averaged across each of the two sequences compared and, thus, subjective impressions from Figs. 6 and 7 may not be confirmed for individual sample frames. Distortion increases for HEVC in going from low QP (high quality) to high QP (low quality), though this change is the combined effect of lower quality video at the higher QPs and the addition of SE. A surprising result, perhaps, is that the gain from HEVC rather than H.264/AVC transrating of encrypted content in terms of reduced bitrate is relatively larger at lower qualities. This is perhaps surprising because HEVC was designed for high resolution and high quality video, which is necessary as visual artifacts are otherwise more apparent at higher resolutions. Comparing between *Stefan* and *Mobile* video sequences, a similar pattern of results is evident. However, there are greater differences in quality between the codecs for *Mobile*, with greater spatial complexity than *Stefan*. It is more apparent from the SSIM index results in Fig. 8 that at QP level 12, SE by the given method is unsuitable when an HEVC codec is used because there is insufficient distortion across the sequences. For spatially more complex *Mobile* as well, if SE is followed by transrating, HEVC processing results in reduced distortion. Therefore, there is a content-dependent and quality-dependent effect when moving between H.264/AVC to HEVC when this type of processing takes place.

**Table 3** Crypto-Transrating results for the HEVC and H.264/AVC video codec with the *Stefan* video sequence

| QP | Bit Rate (AVC) | Bit Rate (HEVC) | Δ Bit Rate (Avg. bit rate) | PSNR_Y(dB) AVC | PSNR_Y(dB) HEVC | Δ PSNR (Y) (dB) | Δ Enc. Time (ms) |
|---|---|---|---|---|---|---|---|
| 12 | 1670 kb | 1130 kb | - 32.34 % | 10.01 | 16.19 | +61.74 % | +81.79 % |
| 24 | 564 kb | 375 kb | -33.51 % | 10.65 | 14.03 | +31.74 % | +33.09 % |
| 36 | 204 kb | 114 kb | -44.12 % | 10.48 | 12.83 | +22.42 % | +14.93 % |
| 48 | 159 kb | 29 kb | -81.76 % | 11.02 | 12.58 | +14.16 % | +9.18 % |

**Table 4** Crypto-Transrating results for the HEVC and H.264/AVC video codec with the *Mobile* video sequence

| QP | Bit Rate (AVC) | Bit Rate (HEVC) | Δ Bit Rate (Avg. Bit-Rate) | PSNR_Y(dB) AVC | PSNR_Y(dB) HEVC | Δ PSNR (Y) (dB) | Δ Enc. Time (ms) |
|---|---|---|---|---|---|---|---|
| 12 | 1755 kb | 1213 kb | -30.88% | 6.7 | 16.5 | +146.27 % | +113 % |
| 24 | 691 kb | 429 kb | -37.92 % | 6.7 | 16.7 | +149.25 % | +82.63 % |
| 36 | 252 kb | 113 kb | -55.16 % | 6.7 | 12.9 | +92.54 % | +61.20 % |
| 48 | 178 kb | 55 kb | -80.29 % | 7.1 | 9.1 | +28.17 % | +34.29 % |

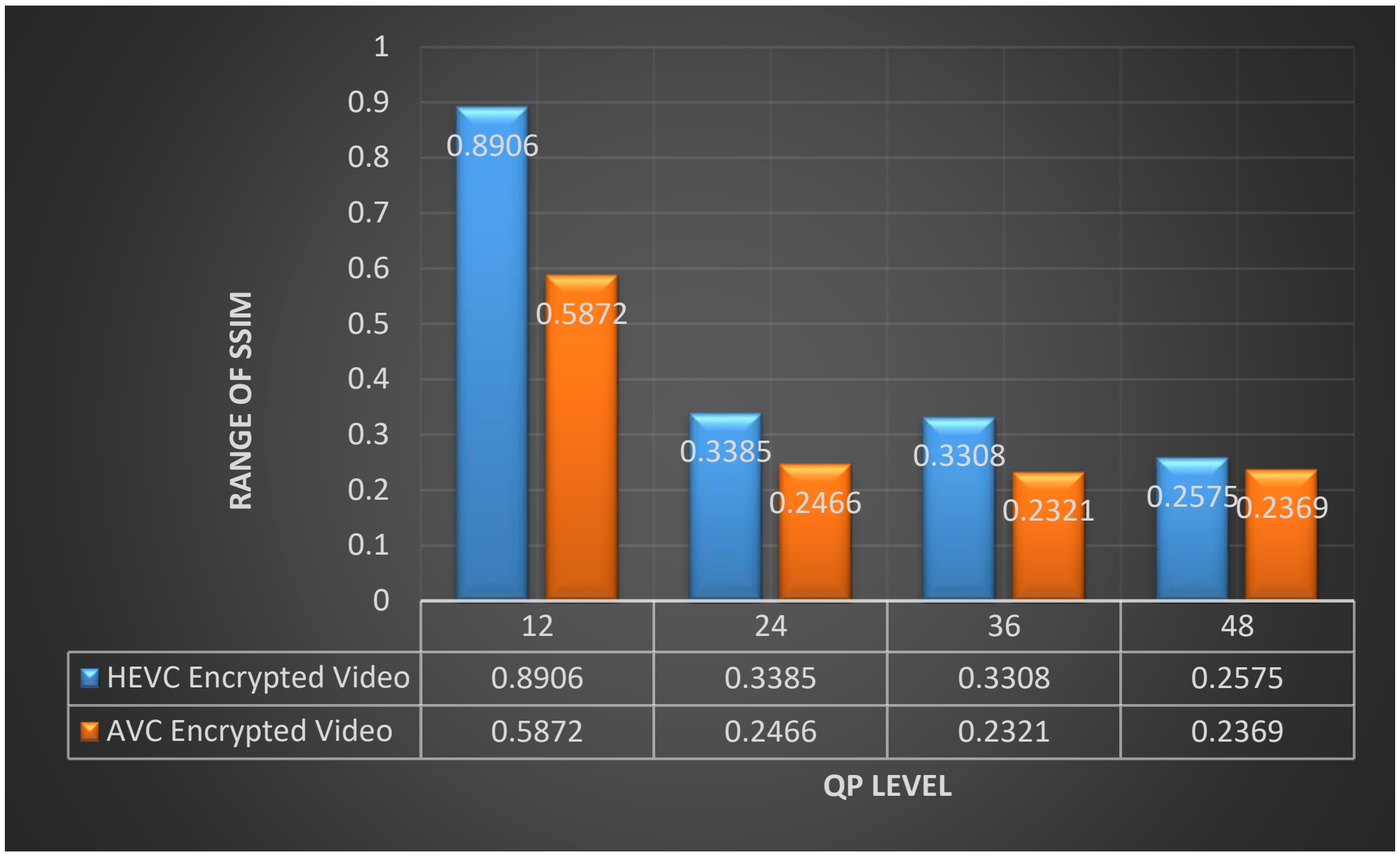

**(a)**

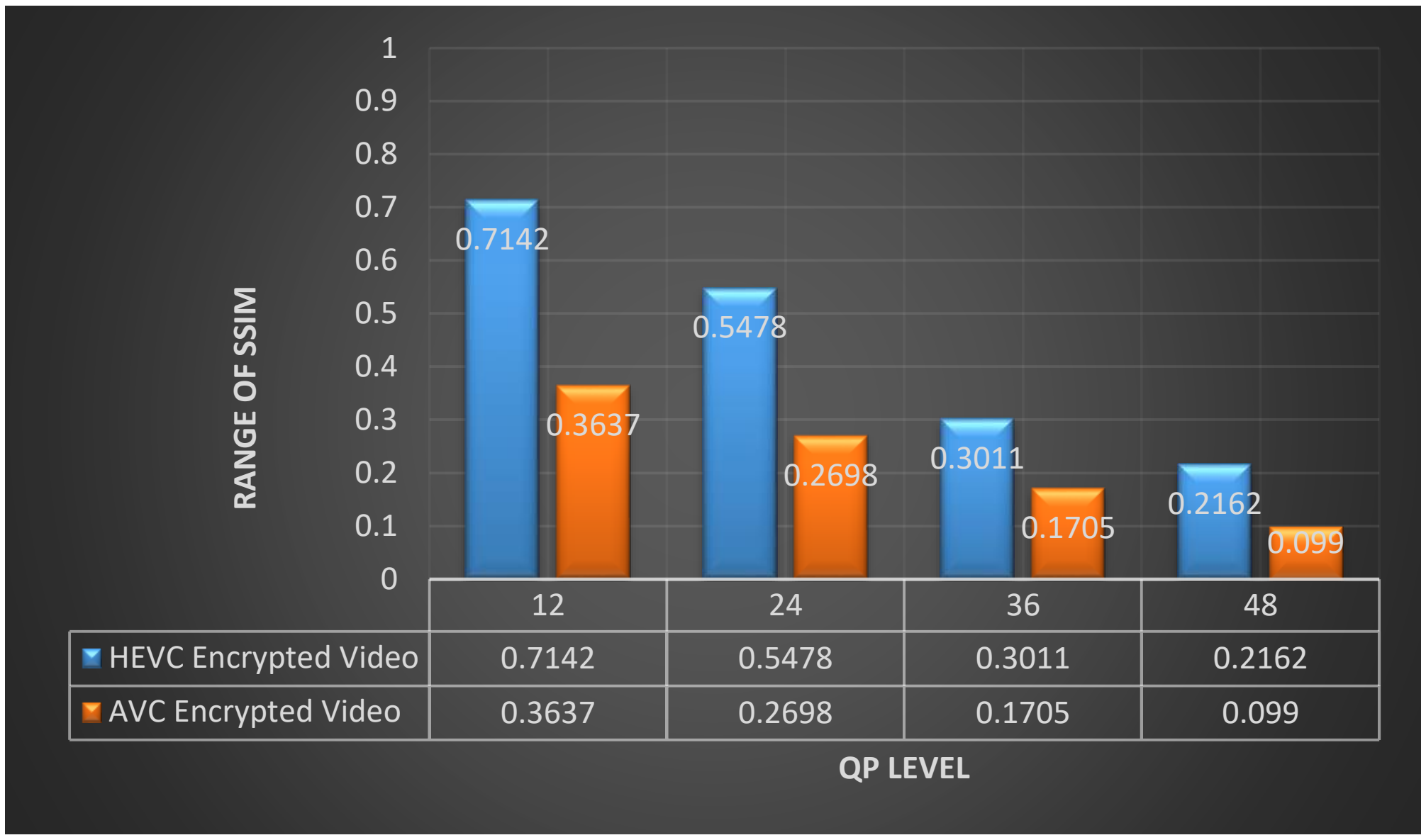

**(b)**

**Fig. 8** Average SSIM values for H.264/AVC and HEVC for a) *Stefan* and b) *Mobile* crypto-transrated videos

## 5.2 HEVC encoded Crypto-transrating with XOR and AES ciphers

The performance of crypto-transcoding with an XOR cipher is tested with H.264/HEVC encoders in Section 5.1 and presented in Figs. 6 and 7. In this Section, transrating after encryption with the state-of-the-art AES cipher is also tested. Fig. 9 shows the visual results of crypto-transrating with the two implemented ciphers i.e. XOR and AES, operating over the same syntax elements and encoded with the HEVC encoder on the HD720 *Four People* video. AES has many modes of operation. In the Cipher Feedback mode (CFB), AES works as a stream-cipher [28] and additionally benefits from a self-synchronization mechanism for real-time transmission. The visual results (from frame no. 53) indicate that the encryption performed using AES-CFB produces a strongly distorted frame, whereas applying an XOR operation results in insufficient distortion. Histogram analysis applied to crypto-transrated video with the two said ciphers is presented in Fig. 10. Fig. 10 contains histograms of the red, green, blue and luminance values of each pixel of the tested video sequence. This Fig. 10 (c, e) indicates that the AES-CFB cipher results in an increase in entropy or randomness across the pixel values relative to using the XOR cipher. This entropy is evident in the spreading of more black and sharp colours across the video frames compared to the original histogram values prior to crypto-transrating of the *Four*

*People* video, Fig.10 (a). This finding implies that, if the videos are selectively encrypted by AES, it is difficult to infer the presence of an object in any one of the R, G, B and luminance domains.

| | Results taken through XOR cipher | Results taken through AES (CFB) |
|---|---|---|
| (a) **Step 1** SE with crypto-encoder at QP 12 | 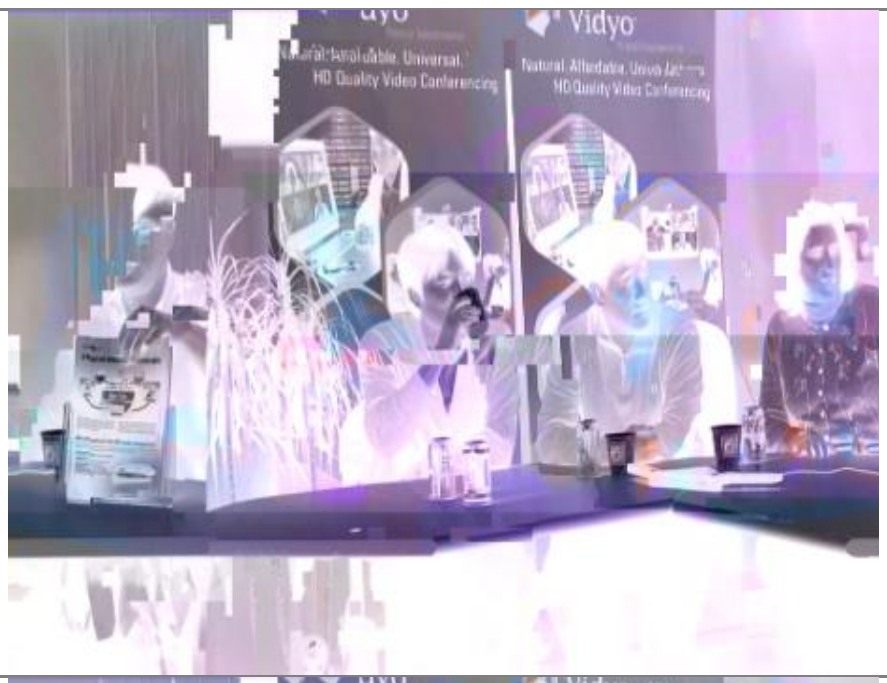 | 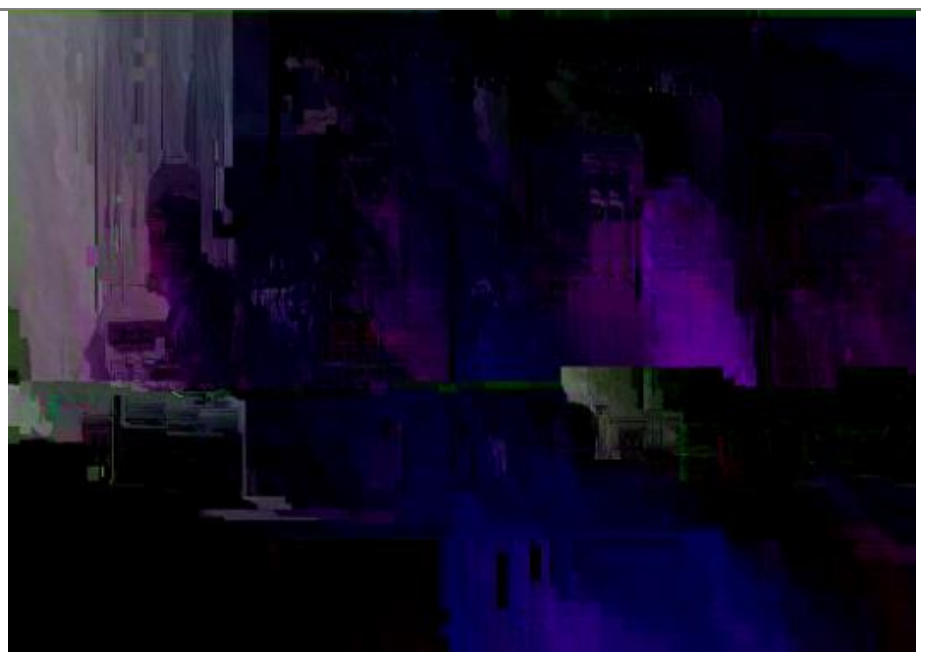 |
| (b) **Step 3** Crypto-encoded video taken from Step 1, transrated with QP 24 | 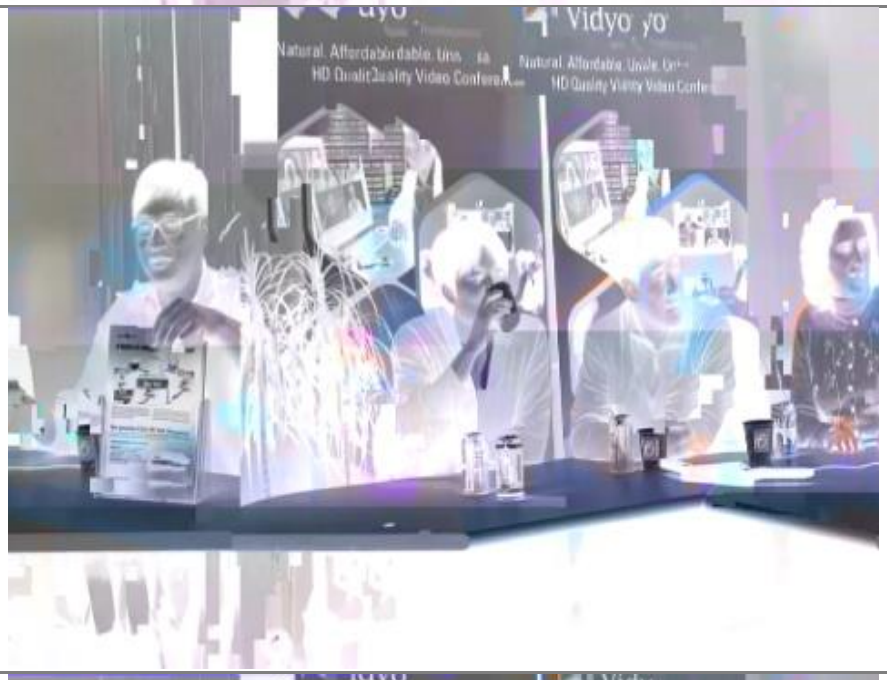 | 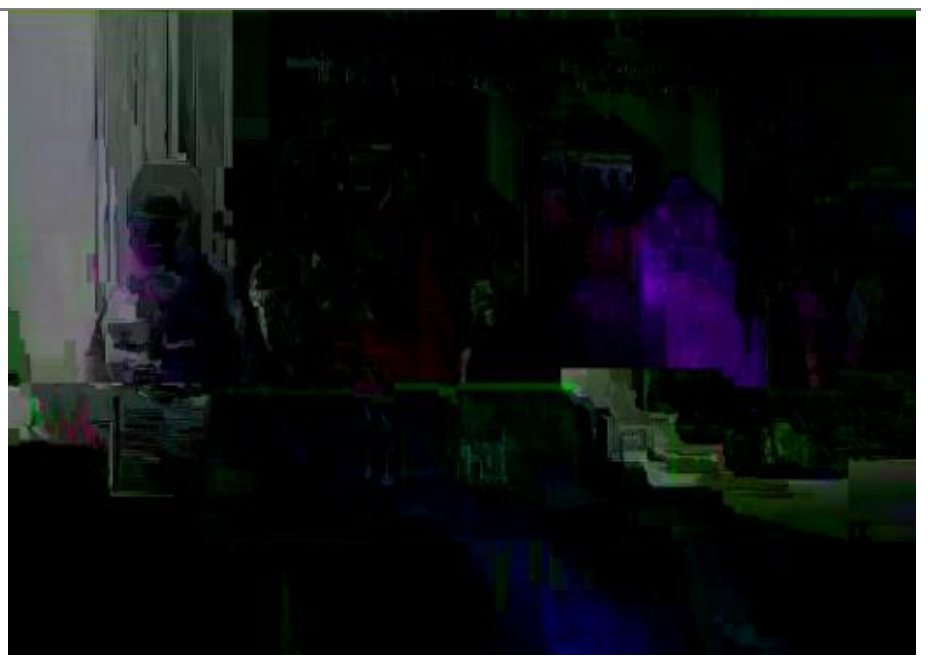 |
| (c) **Step 3** Crypto-encoded video taken from Step 1, transrated with QP 36 | 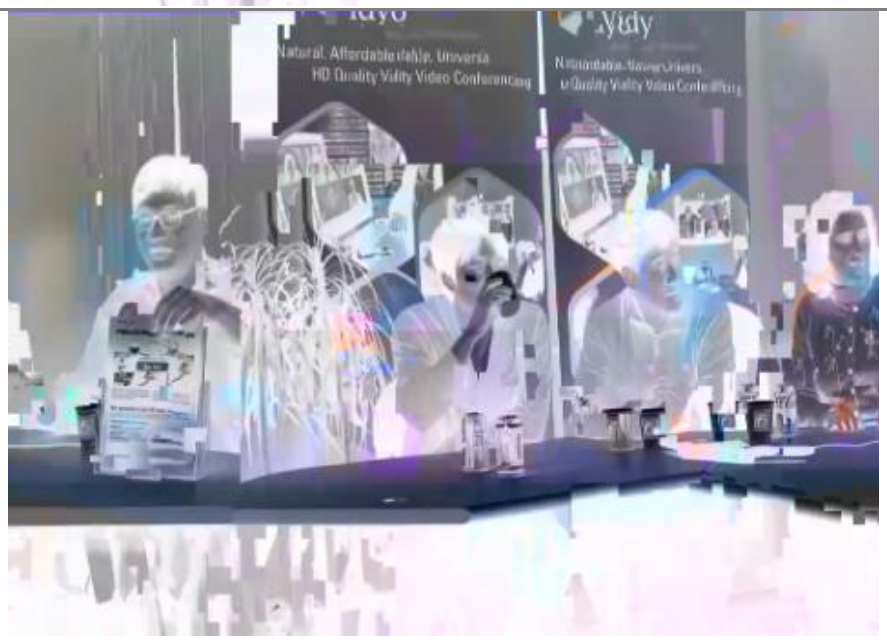 | 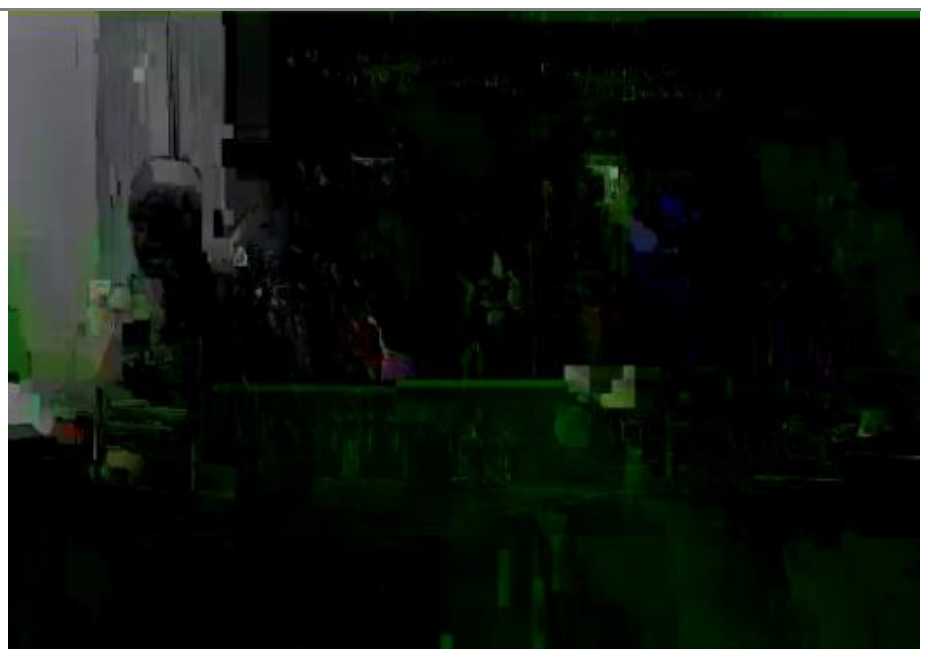 |
| (d) **Step 3** Crypto-encoded video taken from Step 1, transrated with QP 48 | 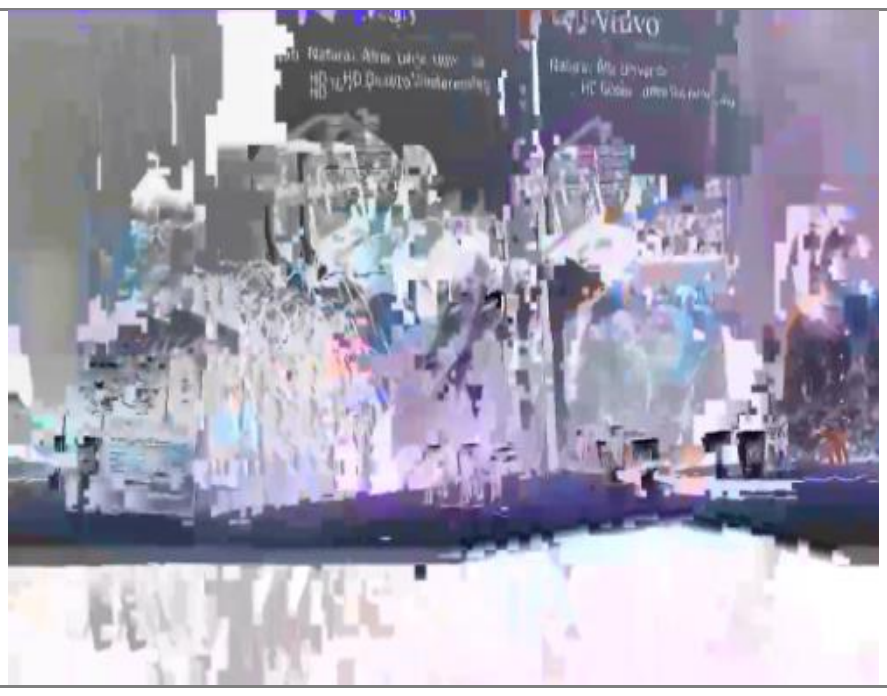 | 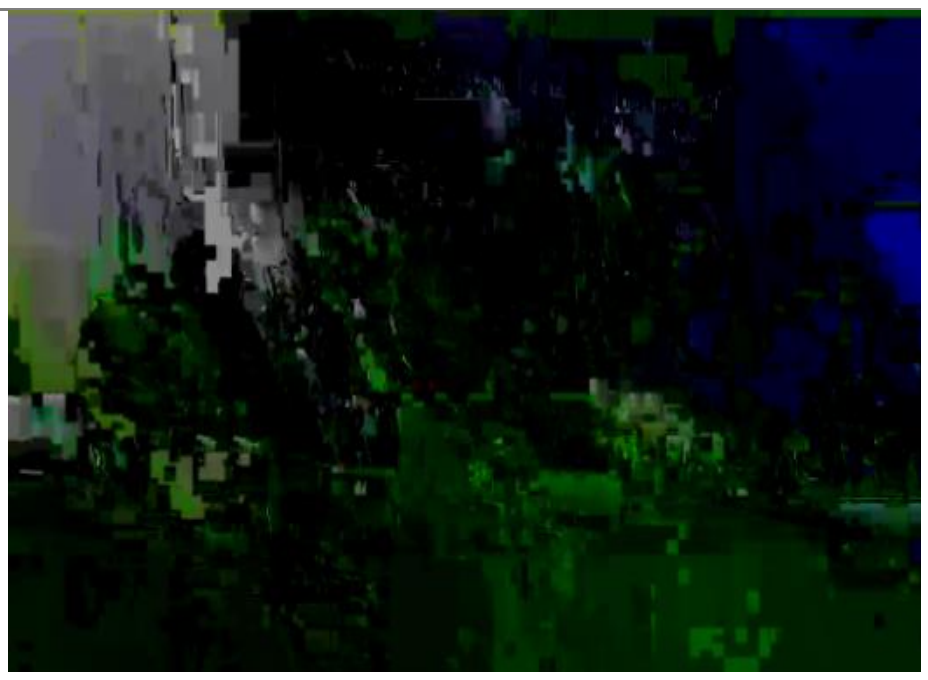 |

**Fig. 9** Visual results for *Four People* (frame #53) after crypto-transrating through XOR and AES ciphers

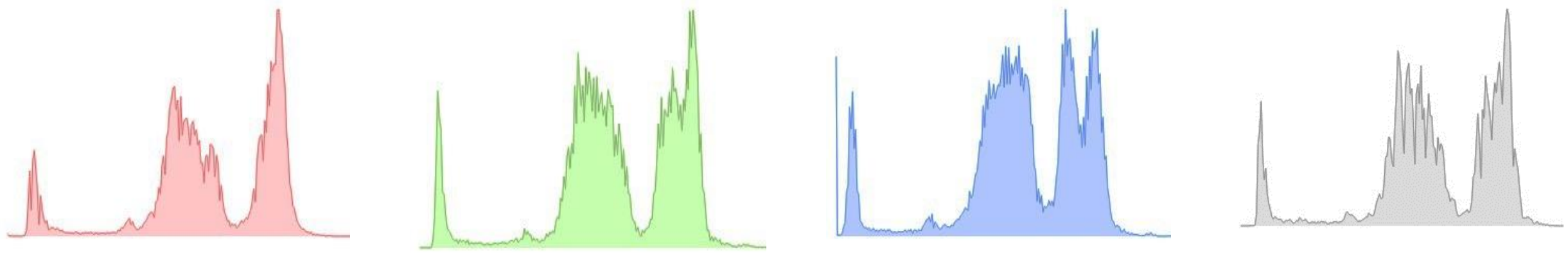

(a) RGB histograms & luminance histogram of original frame

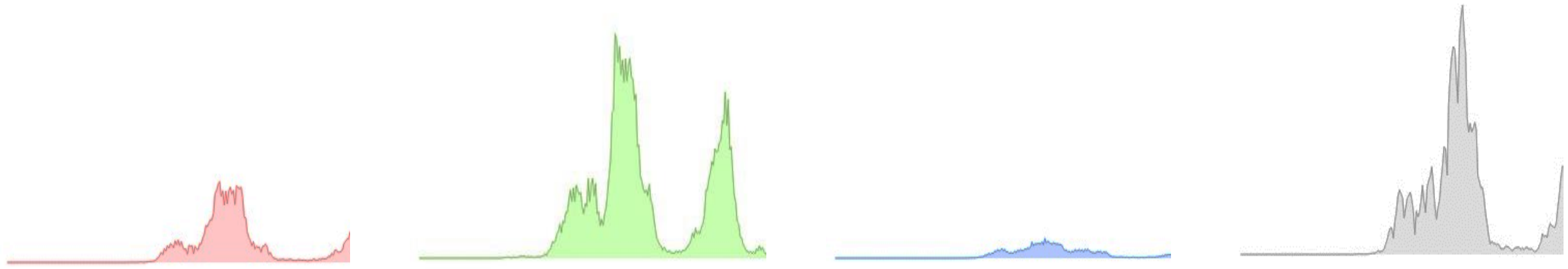

(b) RGB histograms & luminance histograms of video encrypted with XOR at QP 12

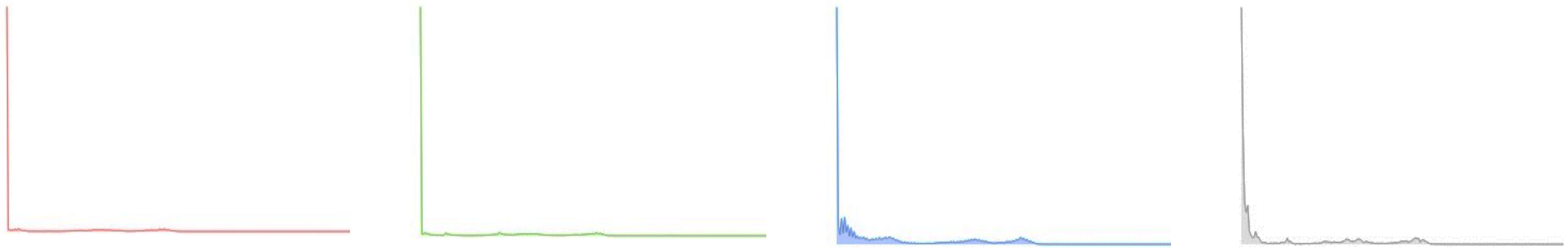

(c) RGB histograms & luminance histogram of video encrypted with AES (CFB) at QP 12

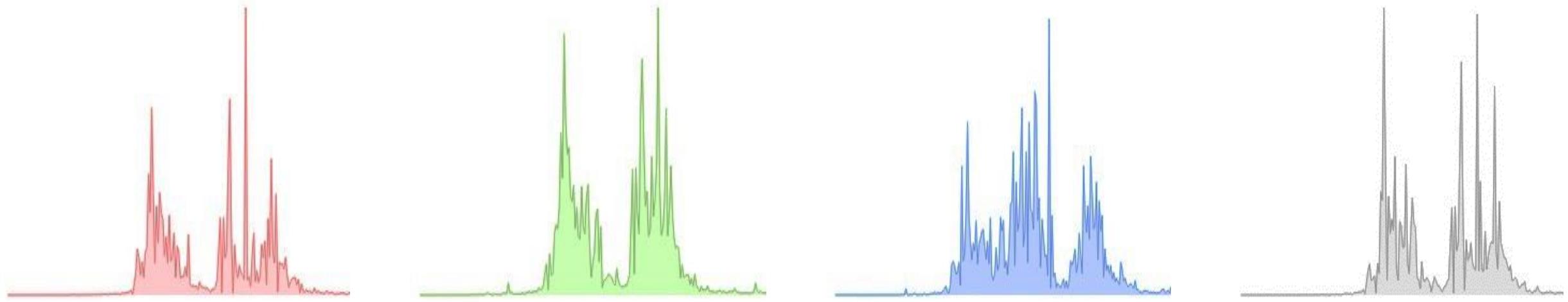

(d) RGB histograms & luminance histograms of video encrypted with XOR and then transrated with QP 48

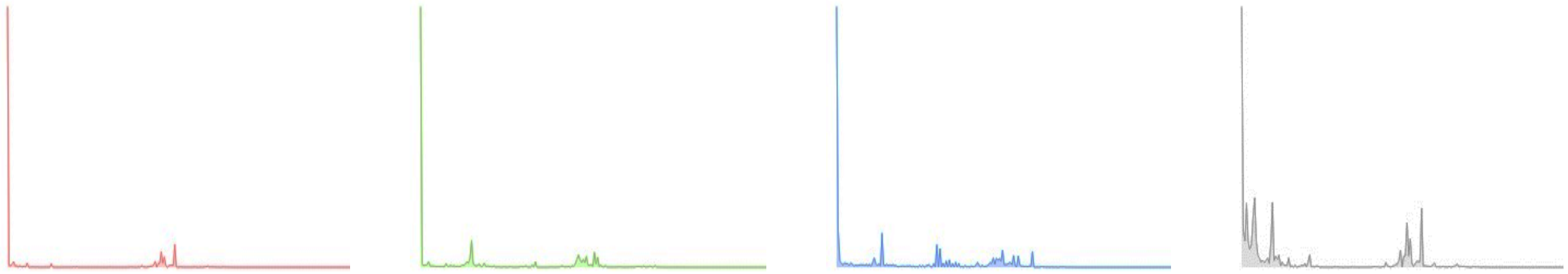

(e) RGB histograms & luminance histograms of video encrypted with AES (CFB) and then transrated with QP 48

**Fig. 10** Histogram analysis of *Four People* video after crypto-transrating with XOR and AES ciphers and encoding with an HEVC encoder

### 5.3 Comparative analysis

Table 5 is an analysis comparing the scheme described with previous secure transcoders. The crypto-transcoder has advantages over previously proposed schemes in [15], [54] and [58], though to some extent these are more to do with the development of the HEVC codec than merits of prior schemes. The proposed crypto-transcoder scheme provides a greater compression rate due to the possible use of an HEVC codec, as well as preserving the whole video structure at transcoders. The proposed scheme also implements real-time transcoding as the authors did in [15] and [54]. Meanwhile, when H.264/AVC is targeted, the computational overhead in terms of encryption with transcoding time appears to be lower than the schemes proposed in [15], [54], and [58].

**Table 5** Comparison with some previous secure transcoding schemes

| Parameters for comparison | (Díaz-Sánchez et al. 2016) [15] | (Boyadjis et al. 2014) [54] | (Thomas et al. 2010) [58] | Proposed crypto-transcoder scheme |
|---|---|---|---|---|
| Video structure preservation at decoder | Yes: Full encryption produced the actual video structure at decoder. | Yes: Preserved the video format with no bit-rate overhead | Yes: Preserved the cipher synchronization with no loss of compression performance at decoder | Yes: The scheme provides a greater compression rate, as well as preserving the video structure at the decoder |
| Real-time transcoding | Yes, by sharing load on multiple machines | Yes | No | Yes |
| Codec standard | H.264/ SVC | H.264/AVC and HEVC | H.264/AVC | Hybrid model: works both for H.264/AVC and HEVC |
| Encryption type | Full encryption: It proposed a distributed encryption and flexible key management, which facilitates content filtering, key extraction and content decryption at the receiver | Selective Encryption was based on symmetric ciphering and managed by AES | Introduced two methods of SE:<br>**1.** On the basis of full I frame encryption; and<br>**2**. On the basis of sign bits of motion vectors and transform coefficients | SE based on the arithmetic signs of motion vectors difference and the signs of texture data (TC) |
| Computational overhead | High due to full encryption. | Low | High in scheme 1.<br>Low in scheme 2 but produced error drift after transcoding | Low for H.264/AVC, medium for HEVC due to video encoding time |

# 6 Conclusions

This paper implemented a joint crypto-transcoder with two widely deployed video codecs, H.264/AVC and HEVC. The main contribution has been to reduce the processing latency at intermediate transcoders arising from a need to encrypt video for reasons of content protection and possibly privacy protection at the user. The paper proposes that a suitable selective-encryption scheme is applied in that the encrypted video bitstream remains decoder format compatible. Because the selective encryption on carefully selected bin-strings reduces the computational overhead of encryption, there is a further gain, apart from the desired reduction in latency, in terms of an overall bitrate reduction. The experiments performed have shown that transcoding of HEVC video produces better results in terms of reduced file sizes. However, if one considers the computation cost of performing joint crypto-transcoding, then HEVC results in greater transcoding times. Therefore, if transcoding is likely to be used then it is better to use SE rather than full encryption, unless the application is (say) military, legal, or medical. From experiments, the gains from using SE are substantial in terms of limited bitrate overhead and (according to the machine employed) several seconds saved in encoding alone, even for short video sequences. For longer video streams, including broadcast TV and films, the savings in bit-rate and computation time will be considerable. If reduced bandwidth is not a priority then the H.264/AVC codec remains viable, unless a hardware HEVC codec is available. From comparisons of visual distortion in the paper, it is apparent that there is a content dependency and QP-dependency when transrating to lower bitrates, in the sense that SE may contribute reduced distortion at higher bitrates. In going from H.264/AVC to HEVC, it is also likely that there will be less distortion at the same QP. In fact, it may be better to avoid HEVC joint crypto-transcoding at the lowest QPs, i.e. for broadcast-quality video. The paper also considers whether a low-complexity cypher is worth considering because of a further reduction in latency at the server and client ends. However, from cipher comparisons between AES and XOR encryption of the selected elements in the video stream, XOR appears insufficient, despite interest in lightweight encryption using XOR for smart grid applications.

Future work will consider how to select suitable syntax elements according to the type of content and the expected transcoded quality. The aim will be to distort those features of a video frame that some encrypted syntax elements do not presently have an impact upon. In that sense, an intelligent SE scheme will adaptively apply encryption to elements within the compressed video stream. At the same time, any such system should preserve decoder format compatibility and minimize any increase in bitrate. There is also scope for performing a number of other statistical tests to confirm the results, such as through correlation-coefficient analysis as an alternative to histogram analysis, and finding the video encryption quantity as an alternative way to investigate the QP-dependency of encryption. Rate-distortion analysis, such as through the well-known Bjøntegaard-Delta metric, is an alternative way of examining the bitrate overhead from encryption.

## Author Bios

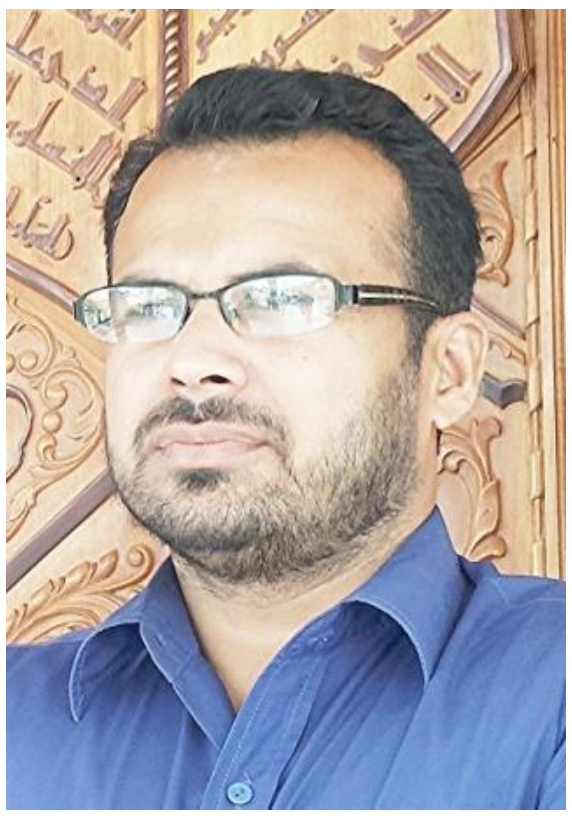

Rizwan Ali Shah received his Master degree of Computer Science from the Virtual University of Pakistan and the M.Phil degree in Computer Science from the Department of Computer Science & IT (DCS&IT), The Islamia University of Bahawalpur (IUB), Pakistan. Currently, he is a PhD Student with DCS & IT, IUB. He is an active Research member of Multimedia Research Group in DCS & IT, IUB. He is working as a Computer Instructor in Federal Government Educational Institutes (Cantt. /Garrison). He has more than 6 years of teaching and R&D experience. He has published an International Conference paper. His research interests are the security aspects of multimedia (audio and video), compression, encryption, secure transcoding and secure multimedia transmission.

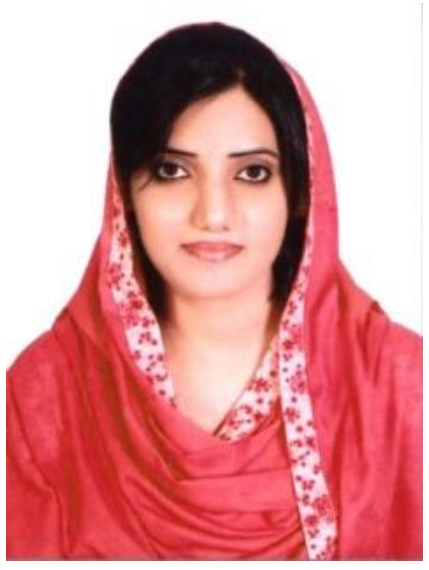

Mamoona N. Asghar received the Bachelors degree in Computer Science from the Islamia University of Bahawalpur (IUB), Punjab, Pakistan, and the Masters degree in Computer Science with the major in Computer Networks Security, from International Islamic University, Islamabad, Pakistan. Her PhD degree was with the School of Computer Science and Electronic Engineering, University of Essex, Colchester, UK in 2013. Currently, she is working as Marie Sklodowska Curie Career-Fit Research Fellow in Software Research Institute, Athlone Institute of Technology (AIT), Ireland since June 2018. She is also a regular faculty member in IUB and worked as an Assistant Professor with the Department of Computer Science and Information Technology (DCS & IT), IUB, Punjab, Pakistan before joining AIT. She has more than 14 years of teaching and R&D experience. She has published several ISI indexed journal articles with numerous International conference papers. She is also working as reviewer of renowned journals (including IEEE Transactions of Dependable and Secure Computing and IEEE Transactions on Information Forensics and Security) and conferences. She has received the Competitive Research Grant awarded by the Higher Education Commission (HEC) Pakistan under National Research Program for Universities (NRPU) 2016. Her research interests include security aspects of multimedia (image, audio and video), compression, encryption, steganography, secure transmission in future networks, video quality metrics, and key management schemes.

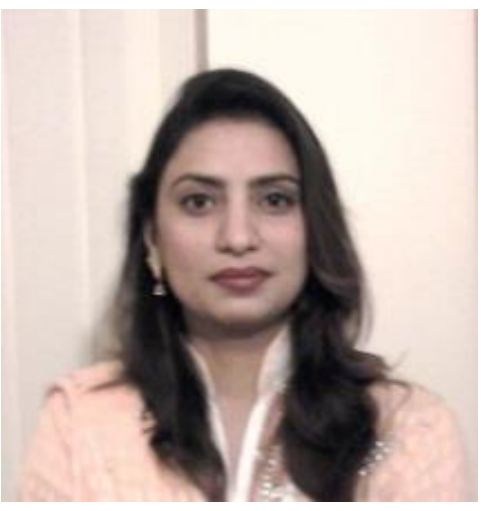

SAIMA ABDULLAH received her Ph.D. from the Department of Computer Science and Electronic Engineering of University of Essex, UK. She is currently working as an Assistant professor at the Department of Computer Science & Information Technology, The Islamia University of Bahawalpur, Pakistan. Her main research interests include wireless networks/communications, future Internet technology and network performance analysis. She has published around 10 papers in the above research areas. She serves as a reviewer of international journals. She is the Member of Multimedia Research Group in DCS & IT, and has been working on efficient and secure communication of multimedia data over future generation network technologies.

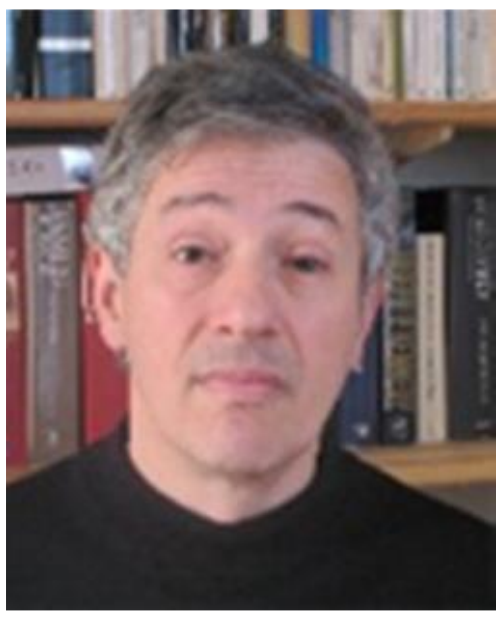

Martin Fleury holds a degree in Modern History (Oxford University, UK) and a Maths/Physics-based degree from the Open University, Milton Keynes, UK. He obtained an MSc in Astrophysics from QMW College, University of London, UK in 1990 and an MSc from the University of South-West England, Bristol in Parallel Computing Systems in 1991. He gained a PhD in Parallel Image-Processing Systems from the University of Essex, Colchester, UK. He worked as a Senior Lecturer at the University of Essex, where he is still a Visiting Fellow. He is now a free-lance consultant. Martin has authored or co-authored around two hundred and fifty-five articles and book chapters on topics such as document and image compression algorithms, performance prediction of parallel systems, software engineering, reconfigurable hardware and vision systems. His current research interests are video communication over wireless networks. He has published or edited books on high performance computing for image processing and peer-to-peer streaming.

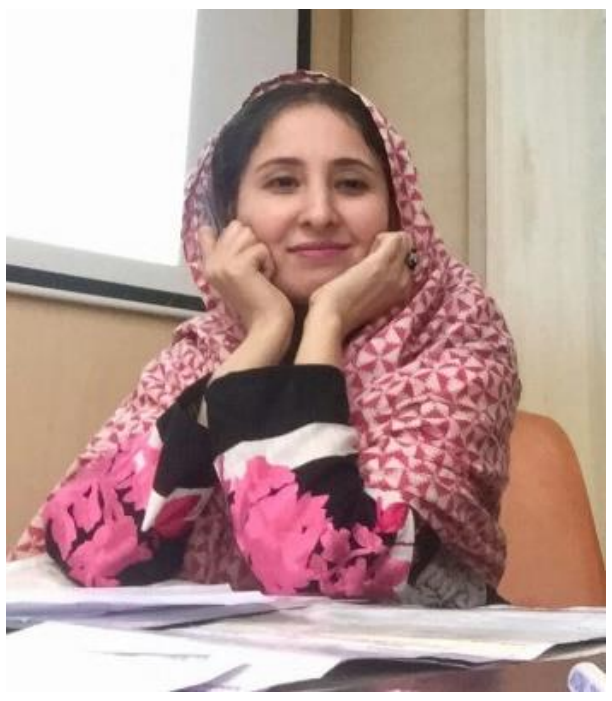

Neelam Gohar is Assistant Professor, Coordinator Advanced Studies and In-charge of Department of Computer Science, Shaheed Benazir Bhutto Women University Peshawar. She completed her PhD from University of Liverpool, UK in 2012. Her research areas are security aspects in Artificial Intelligence technologies, multi-agent decision problems, Computational Social Choice theory and voting system.